\documentclass[%
reprint,
amsmath,amssymb,
]{revtex4-2}

\usepackage{graphicx}
\usepackage{dcolumn}
\usepackage{bm}
\usepackage{braket}
\usepackage{hyperref}
\usepackage{amsmath}
\DeclareMathOperator{\sinc}{sinc}
\usepackage{bm}
\usepackage{float}
\usepackage{comment}

\begin{document}
	
	\preprint{APS/123-QED}
	
	\title {Structured position-momentum entangled two-photon fields}

	\author{Radhika Prasad}
	\email{radhikap@iitk.ac.in}
	\author{Sanjana Wanare}
	\altaffiliation[Currently at ]{Boston University, 8 St. Mary's St., Boston 02215, USA}
	\author{Suman Karan}
	\author{Mritunjay K. Joshi}
	\altaffiliation[Currently at ]{Eberly College of Science, Penn State University, University Park, PA 16802, USA}
	\author{Abhinandan Bhattacharjee}
	\altaffiliation[Currently at ]{Universität Paderborn, Warburger Strasse 100, 33098 Paderborn, Germany}
	\author{Anand K. Jha}
	\email{akjha@iitk.ac.in}
	\affiliation{Department of Physics, Indian Institute of Technology Kanpur, Kanpur, UP 208016, India}
	
\date\today	
	
	
	\begin{abstract}
		Structured optical fields have led to several ground-breaking techniques in classical imaging and microscopy. At the same time, in the quantum domain, position-momentum entangled photon fields have been shown to have several unique features that can lead to beyond-classical  imaging and microscopy capabilities. Therefore, it is natural to expect that position-momentum entangled two-photon fields that are structured can push the boundaries of quantum imaging and microscopy even further beyond. Nonetheless, the existing experimental schemes are able to produce either structured two-photon fields without position-momentum entanglement, or position-momentum entangled two-photon fields without structures. In this article, by manipulating the phase-matching condition of the spontaneous parametric down-conversion process, we report experimental generation of two-photon fields with various structures in their spatial correlations. We experimentally measure the minimum bound on the entanglement of formation and thereby verify the position-momentum entanglement of the structured two-photon field. We expect this work to have important implications for quantum technologies related to imaging and sensing.

	\end{abstract}
	
	\maketitle
	
	
\section{\label{sec:1}Introduction}
	
Structured light refers to shaping of light fields by manipulating its amplitude, phase, polarisation, and correlation function. Structured light fields have several applications including the ground-breaking imaging and microscopy techniques such as stimulated emission depletion microscopy (STED) \cite{hell1994optlett} and super resolution microscopy (SIM) \cite{samanta2021jopt}, super-resolution imaging with structured point spread function \cite{curcio2020natcomm}, detection of transverse motion of particles \cite{rosales2013scirep}, Overcoming the classical Rayleigh diffraction limit by controlling two-point correlations of partially coherent light sources \cite{liang2017optexp}, self healing of optical fields \cite{shen2022jopt}, and optical communication \cite{bozinovic2013science, wang2012science}. For a broader overview of structured light fields, see  Refs. \cite{forbes2021natphot,rubinsztein2016jopt,he2022lsciapp,shen2022jopt}.

On the other hand, in the last three decades, quantum entanglement has emerged as a resource for performing tasks that would otherwise be impossible. Several quantum communication \cite{dambrosio2012natcomm, marcikic2003nature, ecker2019prx} and quantum cryptography \cite{ekert1991prl,jennewein2000prl, liao2017natcomm} applications based on entanglement have been proposed and demonstrated. The other major application of quantum entanglement has been in quantum imaging and microscopy. These include ghost imaging \cite{strekalov1995prl, pittman1995pra, moreau2018ghost},  sub-shot-noise imaging \cite{brida2010natphot,blanchet2008prl}, resolution-enhanced imaging \cite{giovannetti2009pra,unternahrer2018optica,santos2022prl}, quantum microscopy of cell at the Heisenberg limit \cite{he2023natcomm}, quantum imaging with undetected photons \cite{lemos2014nature}, quantum image distillation \cite{fuenzalida2023sciadv, defienne2019sciadv}, adaptive optical imaging with entangled photons \cite{cameron2024science}, quantum holography \cite{defienne2021natphy}, and information encoding using entangled twin beams \cite{nirala2023sciadv}. For a more comprehensive review of such applications, see Refs.   \cite{genovese2016jopt,moreau2019natrevphy,erkmen2010advoptphot,padgett2017philtransrsa}. 
Most of these quantum imaging schemes use the entangled  two-photon field  produced by spontaneous parametric down-conversion (SPDC) in a nonlinear crystal. \cite{karan2020jopt, hong1985pra}. Entanglement in a given degree of freedom implies simultaneous strong correlations in two conjugate bases. Although the SPDC photons can become entangled in various degrees of freedom including polarisation \cite{kwiat1995prl}, angle-orbital angular momentum (OAM) \cite{leach2010science}, time-frequency \cite{thew2004prl} and position-momentum \cite{o2005prl,walborn2010phyrep}, the quantum imaging and microscopy applications have mostly been based on position-momentum entanglement.

\begin{figure*}[!t]
	\centering
	\includegraphics[scale=0.85]{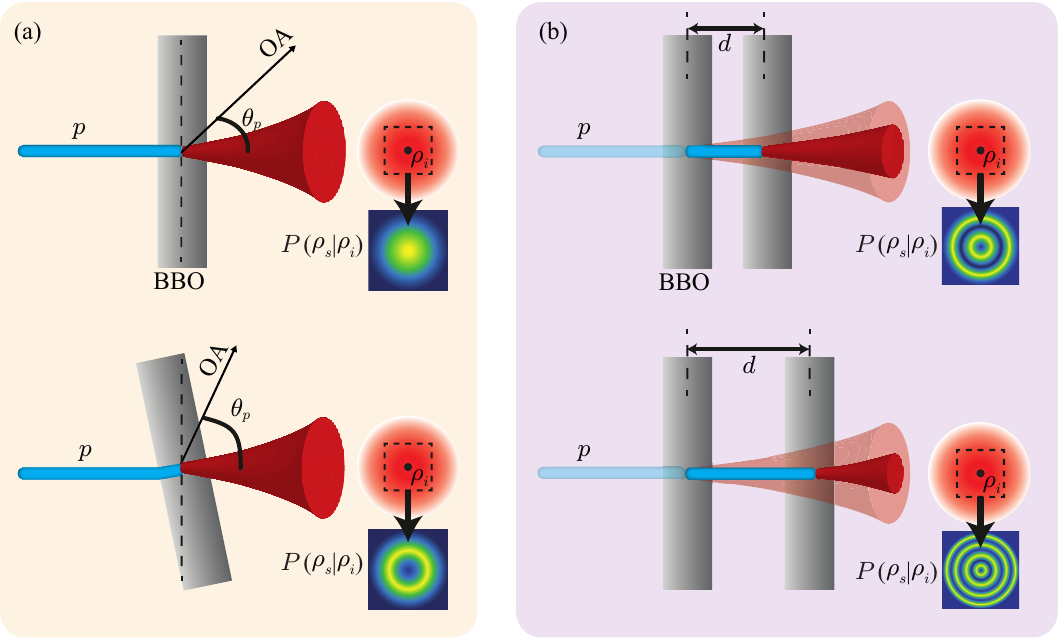}
	\caption{Two methods for manipulation the phase-matching function. (a) by changing the phase matching angle $\theta_p$ of a single crystal. (b) by generating SPDC from a pair of BBO crystals separated by a distance $d$. OA stands for optic axis.}
	\label{fig:phase matching}
\end{figure*}

The position-momentum entanglement implies strong correlation in both the position and momentum of the two photons. One way to estimate the correlation in a given variable is conditional uncertainty, which refers to the uncertainty of individual photons given that the other photon has been detected \cite{bhattacharjee2022njp}. An experimental violation of Heisenberg uncertainty relation by the conditional position and momentum uncertainties is routinely used as a technique for verifying position-momentum entanglement \cite{berta2010natphy, howell2004prl, bhattacharjee2022sciadv, bhattacharjee2022njp}. In the existing quantum imaging techniques \cite{strekalov1995prl, pittman1995pra, moreau2018ghost, brida2010natphot,blanchet2008prl, giovannetti2009pra,unternahrer2018optica,santos2022prl, he2023natcomm, lemos2014nature, fuenzalida2023sciadv, defienne2019sciadv, cameron2024science, defienne2021natphy, nirala2023sciadv}, the structured position-momentum entangled fields have not been employed. However, given the ground-breaking capabilities provided by structured light fields for classical imaging, it is natural to expect that the position-momentum entangled two-photon fields that are structured can push the boundaries of quantum imaging and microscopy even further beyond.

The current methods of generating two-photon fields correlated in position and momentum variables can be divided into two categories.  
The first category consists of methods that generate two-photon position-momentum entanglement field but with no structure \cite{berta2010natphy, howell2004prl, bhattacharjee2022sciadv, bhattacharjee2022njp}. The second category consists of methods that generate two-photon fields with structured correlations without position-momentum entanglement \cite{boucher2021optlett,valencia2007prl,monken1998pra,zhang2019optexp, zia2023natphot}. One way to structure the correlation is by shaping the pump field producing SPDC, and this way of structuring the correlation has quite often been employed in far-field regimes \cite{karan2020jopt}. However, in the far field, the photons produced by SPDC do not remain position-momentum entangled, as witnessed by EPR correlations or entanglement of formation \cite{berta2010natphy, howell2004prl, bhattacharjee2022sciadv}. On the other hand, the phase matching plays the dominant role in shaping the correlation function in the near field where the SPDC photons do remain entangled in the position-momentum degree of freedom. Therefore, in this article, by manipulating the phase matching function, we report experimental generation of structured position-momentum entangled fields in the near-field of the SPDC crystal.

\begin{figure}[t!]
\centering
\includegraphics[scale=0.70]{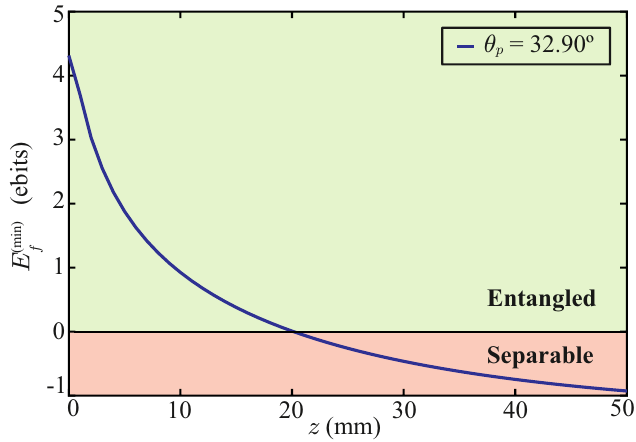}
\caption{The plot of the numerically simulated $E_f^{(\rm min)}$ as a function of $z$ at $\theta_p=32.9^{\circ}$.}
\label{min-EOF}
\end{figure}

\section{Theory}

\subsection{Two-photon probability distribution functions}

\begin{figure*}[!t]
\centering
\includegraphics[scale=0.8]{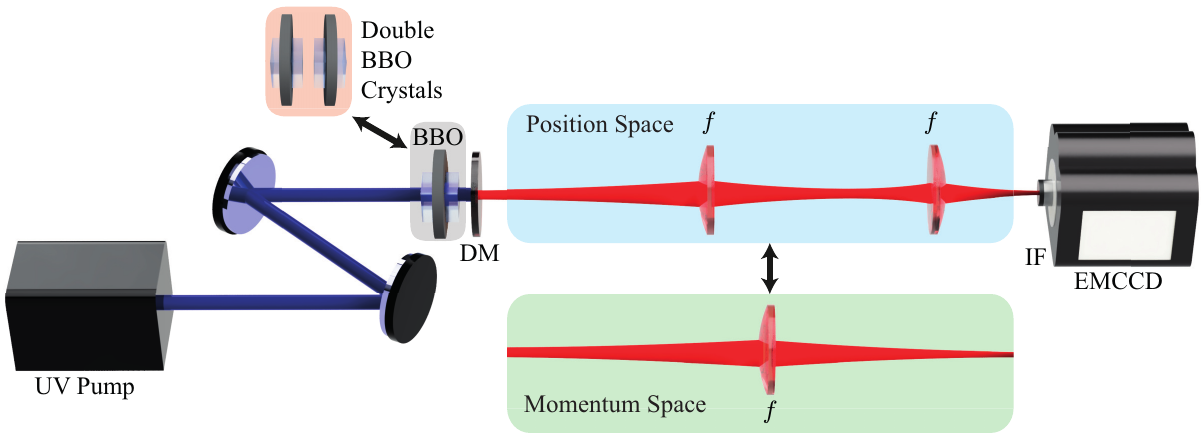}
\caption{Schematic of the experimental setup. UV pump: Ultraviolet laser; DM: Dichroic mirror to block UV pump of 355 nm; IF: Interference filter of 710 nm with 10 nm bandwidth; EMCCD: Electron Multiplying Charge-Coupled Device camera}
\label{fig:exp set up}
\end{figure*}
SPDC is a nonlinear optical process, in which a pump photon of higher frequency gets down-converted into two lower-frequency photons called the signal and idler. Denoting pump, signal, and idler by subscripts $p$, $s$, and $i$, respectively, we write the two-photon state $\left|\Psi\right\rangle$ produced by the SPDC crystal placed at $z=0$ as \cite{karan2020jopt}:
\begin{align}\label{eq:two-photon state}
\!\!\left|\Psi\right\rangle=\iint d^{2} \boldsymbol{q}_{s} d^{2} \boldsymbol{q}_{i} V\left(\boldsymbol{q}_{s}+\boldsymbol{q}_{i}\right) \Phi\left(\boldsymbol{q}_{s},\boldsymbol{q}_{i}\right)\left|\boldsymbol{q}_{s}\right\rangle\left|\boldsymbol{q}_i\right\rangle.
\end{align}
Here $\bm{q}_s$ and $\bm{q}_i$ are the transverse momenta of signal and idler photons, $V\left(\boldsymbol{q}_{s}+\boldsymbol{q}_{i}\right)$ is the pump field amplitude, and $\Phi\left(\boldsymbol{q}_{s},\boldsymbol{q}_{i}\right)$ is the phase-matching function. We take pump to be a Gaussian beam, that is, $V(\boldsymbol{q}_{p})=V_{0}\exp(-\left|\boldsymbol{q}_{p}\right|^{2} w_{0}^{2}/4)$ where $w_0$ is the beam waist and $\bm{q}_p=\bm{q}_s+\bm{q}_i$. Using the two-photon state in Eq.~\ref{eq:two-photon state} and the derivation given in \cite{karan2020jopt}, we obtain the following expression for the two-photon probability distribution function $P_{\rm pos}(\boldsymbol{\rho}_s, \boldsymbol{\rho}_i; z)$ in the transverse position space at a distance $z$ away from the crystal:
\begin{multline}\label{eq:joint probability}
P_{\rm pos}(\boldsymbol{\rho}_s, \boldsymbol{\rho}_i; z)=\Bigg| \iint d^{2} \boldsymbol{q}_{s} d^{2} \boldsymbol{q}_{i}  V\left(\boldsymbol{q}_{s}+\boldsymbol{q}_{i}\right) \Phi\left(\boldsymbol{q}_{s}, \boldsymbol{q}_{i}\right) \\ 
\times\left.\exp \left[i\left(\boldsymbol{q}_{s} \cdot \boldsymbol{\rho}_{s}+\boldsymbol{q}_{i} \cdot \boldsymbol{\rho}_{i}-\frac{\left|\boldsymbol{q}_{s}\right|^{2} z}{2 k_{s}}-\frac{\left|\boldsymbol{q}_{i}\right|^{2} z}{2 k_{i}}\right)\right]\right|^{2}.
\end{multline}
Here $P_{\rm pos}(\boldsymbol{\rho}_s, \boldsymbol{\rho}_i; z)$ is the probability of detecting in coincidence the signal and idler photons at $\boldsymbol{\rho}_s$ and $\boldsymbol{\rho}_i$, respectively. The two-photon conditional probability distribution function $P_{\rm pos}(\boldsymbol{\rho}_s|\boldsymbol{\rho}_{i0}; z)$ of detecting the signal photon, given that the idler photon is detected at $\boldsymbol{\rho}_{i0}$, is obtained by setting $\boldsymbol{\rho}_i=\boldsymbol{\rho}_{i0}$ in Eq.~\ref{eq:joint probability}, that is,
\begin{align}\label{eq:conditional probability}
P_{\rm pos}(\boldsymbol{\rho}_s|\boldsymbol{\rho}_{i0}; z)=P_{\rm pos}(\boldsymbol{\rho}_s, \boldsymbol{\rho}_{i}=\boldsymbol{\rho}_{i0}; z).
\end{align}
For plotting purposes, we work with averaged two-photon probability distribution function $P_{\rm pos}^{(\rm av)}(x_s, x_i; z)$ and the two-photon conditional probability distribution function $P_{\rm pos}(x_s, y_s|0, 0; z)$, defined as:
\begin{align}
P_{\rm pos}^{(\rm av)}(x_s, x_i; z)=\iint P_{\rm pos}(x_s, y_s, x_i, y_i; z)dy_s dy_i, \label{av-joint-pos} \\
P_{\rm pos}(x_s, y_s|0, 0; z)=P_{\rm pos}(x_s, y_s, x_i=0, y_i=0; z). \label{conditional-pos}
\end{align}
Next, using Eq.~(\ref{eq:two-photon state}), we obtain the two-photon probability distribution function in the transverse momentum basis to be $P_{\rm mom}({\bm q}_s, {\bm q}_i; z)=| \langle\boldsymbol{q}_{s}|\langle\boldsymbol{q}_i|\Psi\rangle|^2=|V\left(\boldsymbol{q}_{s}+\boldsymbol{q}_{i}\right) \Phi\left(\boldsymbol{q}_{s},\boldsymbol{q}_{i}\right)|^{2}$. 
Therefore, two-photon conditional probability distribution function $P_{\rm mom}({\bm q}_s, {\bm q}_{i0}; z)$ of detecting the signal photon, given that the idler photon is detected with momentum ${\bm q}_{i0}$ can be written as $P_{\rm mom}({\bm q}_s|{\bm q}_{i0}; z)= P_{\rm mom}({\bm q}_s, {\bm q}_i={\bm q}_{i0}; z)$. We note that the correlation in the momentum basis remains independent of $z$, which is a consequence of the fact that the momentum remains conserved in the SPDC. Finally, we write the averaged two-photon joint probability distribution function as: 
\begin{align}
P_{\rm mom}^{(\rm av)}(q_{xs}, q_{xi}; z)= \iint P_{\rm mom}(q_{xs}, q_{ys}, q_{xi}, q_{yi}; z)dq_{ys} dq_{yi}. \label{av-joint-mom}
\end{align}
\begin{figure*}[!t]
\centering
\includegraphics[scale=0.85]{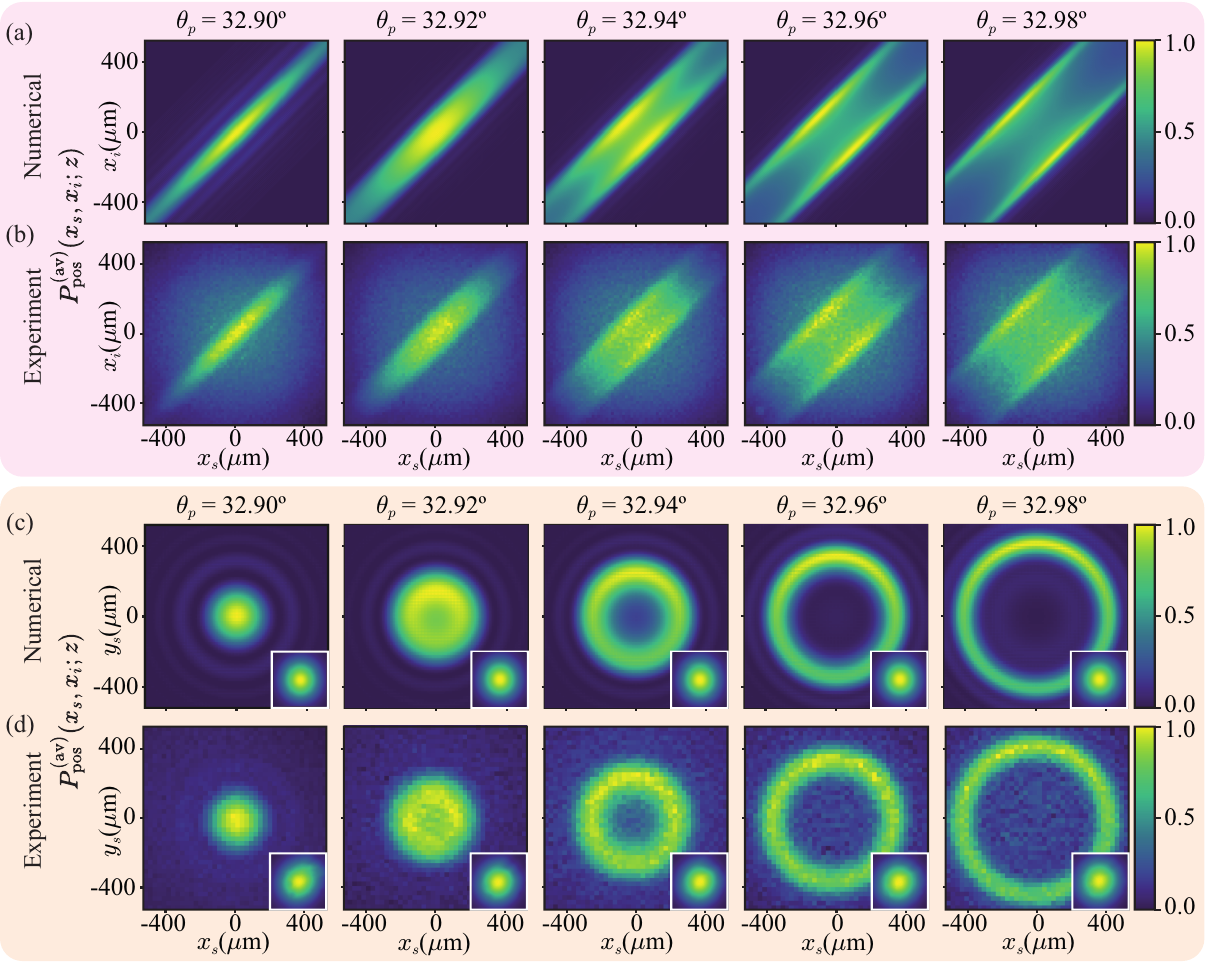}
\caption{(a), (b) Numerically evaluated and experimental measured plots of $P_{\rm pos}^{(\rm av)}(x_s, x_i; z)$ at various $\theta_p$ and at $z=5$ mm. (c), (d) Numerically evaluated and experimental measured plots of $P_{\rm pos}(x_s, y_s|0, 0; z)$ at various $\theta_p$ and at $z=5$ mm.}
\label{fig:1 crystal}
\end{figure*}

\subsection{Structuring correlations by manipulating the phase-matching function}


It is knows that the SPDC photons lose their entanglement in the position-momentum degree of freedom in the far-field regions of the nonlinear crystal \cite{bhattacharjee2022sciadv, howell2004prl}. Therefore, we work in the near-field regions and structure the correlation function by manipulating the phase matching function, which is given by $\Phi\left(\boldsymbol{q}_{s},\boldsymbol{q}_{i}\right) =\int_{0}^{L} \exp(i\Delta k_z z) dz$ \cite{karan2020jopt}. Here, $L$ is the thickness of the crystal and $\Delta k_z$ is called the phase-mismatch parameter and is given by $\Delta k_z = k_{sz} + k_{iz} - k_{pz}$, where $k_{sz}$, $k_{iz}$ and $k_{pz}$ represent the longitudinal wave vectors for signal, idler and pump fields, respectively, and are expressed as \cite{karan2020jopt}:
\begin{align*}
		k_{pz} &= -\alpha_p q_{px} + \eta_p K_{p0} - \frac{ \left[ \beta_p^2 q^2_{px} + \gamma_p^2 q^2_{py}\right]}{2 \eta_p K_{p0}},\\
		k_{sz} &=  n_{so} K_{s0} - \frac{1}{2 n_{so} K_{s0}}(q_{sx}^2+q_{sy}^2), \\
		k_{iz} &=  n_{io} K_{i0} - \frac{1}{2 n_{io} K_{i0}}(q_{ix}^2+q_{iy}^2),
\end{align*}
where $K_{j0}= \frac{2\pi} {\lambda_{j}}$ with  $\lambda_j $ being the wavelength and  $n_{jo}$ and $n_{je}$ being the refractive index for ordinary and extraordinary polarizations. The quantities $\alpha_p$ , $\beta_p$, $\gamma_p$ and $\eta_p$  are given by 
\begin{align*}
\alpha_p = \frac{(n^2_{po}- n^2_{pe})\sin\theta_p \cos\theta_p}{n^2_{po}\sin^2\theta_p + n^2_{pe} \cos^2 \theta_p},\\
		\beta_p = \frac{n_{po} n_{pe}}{n^2_{po}\sin^2\theta_p + n^2_{pe} \cos^2 \theta_p},\\
		\gamma_p = \frac{n_{po} }{\sqrt[]{n^2_{po}\sin^2\theta_p + n^2_{pe} \cos^2 \theta_p}},\\
		\eta_p = \frac{n_{po} n_{pe} }{\sqrt[]{n^2_{po}\sin^2\theta_p + n^2_{pe} \cos^2 \theta_p}}.
\end{align*} 
Here, $\theta_p$ is called the phase-matching angle, which is the angle between the pump propagation direction and the optic axis of the nonlinear crystal. We investigate two different methods of manipulating the phase-matching function and thus of structuring the correlations. The first method is by changing the phase-matching angle $\theta_p$ [see Fig \ref{fig:phase matching}(a)]. In this case, $\Phi\left(\boldsymbol{q}_{s},\boldsymbol{q}_{i}\right)$ can be written as:
\begin{align}\label{eq:single crystal phi}
\Phi\left(\boldsymbol{q}_{s},\boldsymbol{q}_{i}\right)= \sinc\left(\Delta k_z\frac{L}{2}\right)\exp\left(i\Delta k_z\frac{L}{2}\right).
\end{align}
The second method is by having SPDC with two nonlinear crystals separated by distance $d$ [see Fig \ref{fig:phase matching}(b)]. In this case, the generated two-photon state is a superposition of the two-photon states produced by the two crystals. The $\Phi\left(\boldsymbol{q}_{s},\boldsymbol{q}_{i}\right)$ in this case can be shown to be
\begin{align}\label{eq:double crystal phi}
\Phi\left(\boldsymbol{q}_{s},\boldsymbol{q}_{i}\right) = \sinc\left(\Delta k_z\frac{L}{2}\right)\cos\left(\Delta k_z\frac{L+d}{2}\right).
\end{align}
\begin{figure*}[!t]
\centering
\includegraphics[width=2 \columnwidth]{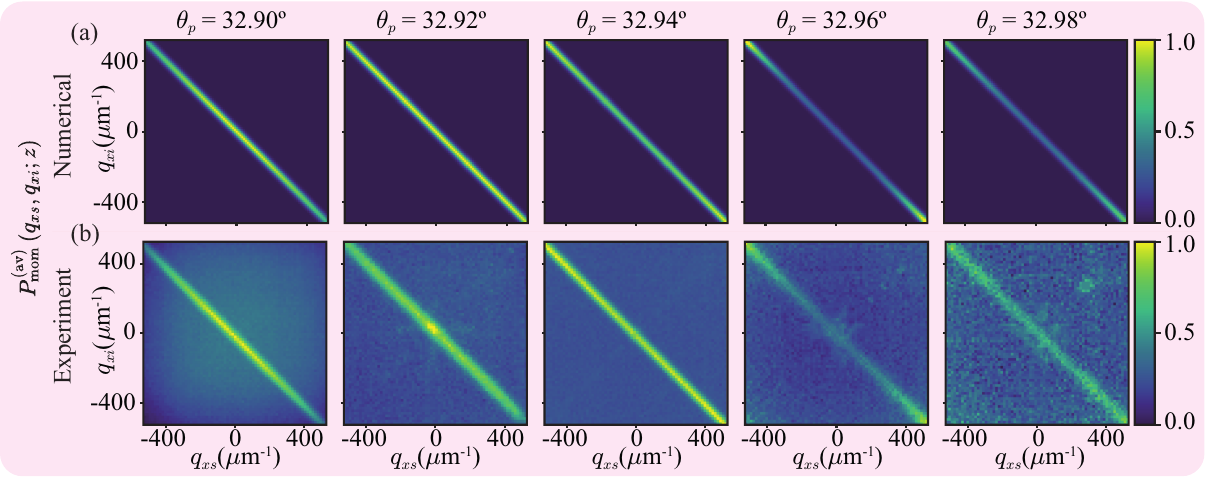}
\caption{(a), (b) Numerically evaluated and experimental measured plots of $P_{\rm mom}^{(\rm av)}(q_{xs}, q_{xi}; z)$ at various $\theta_p$ and at $z=5$ mm.}
\label{fig:mom corr}
\end{figure*}

\subsection{Position-momentum entanglement}
	
\begin{figure*}[!t]
\centering
\includegraphics[scale=0.7]{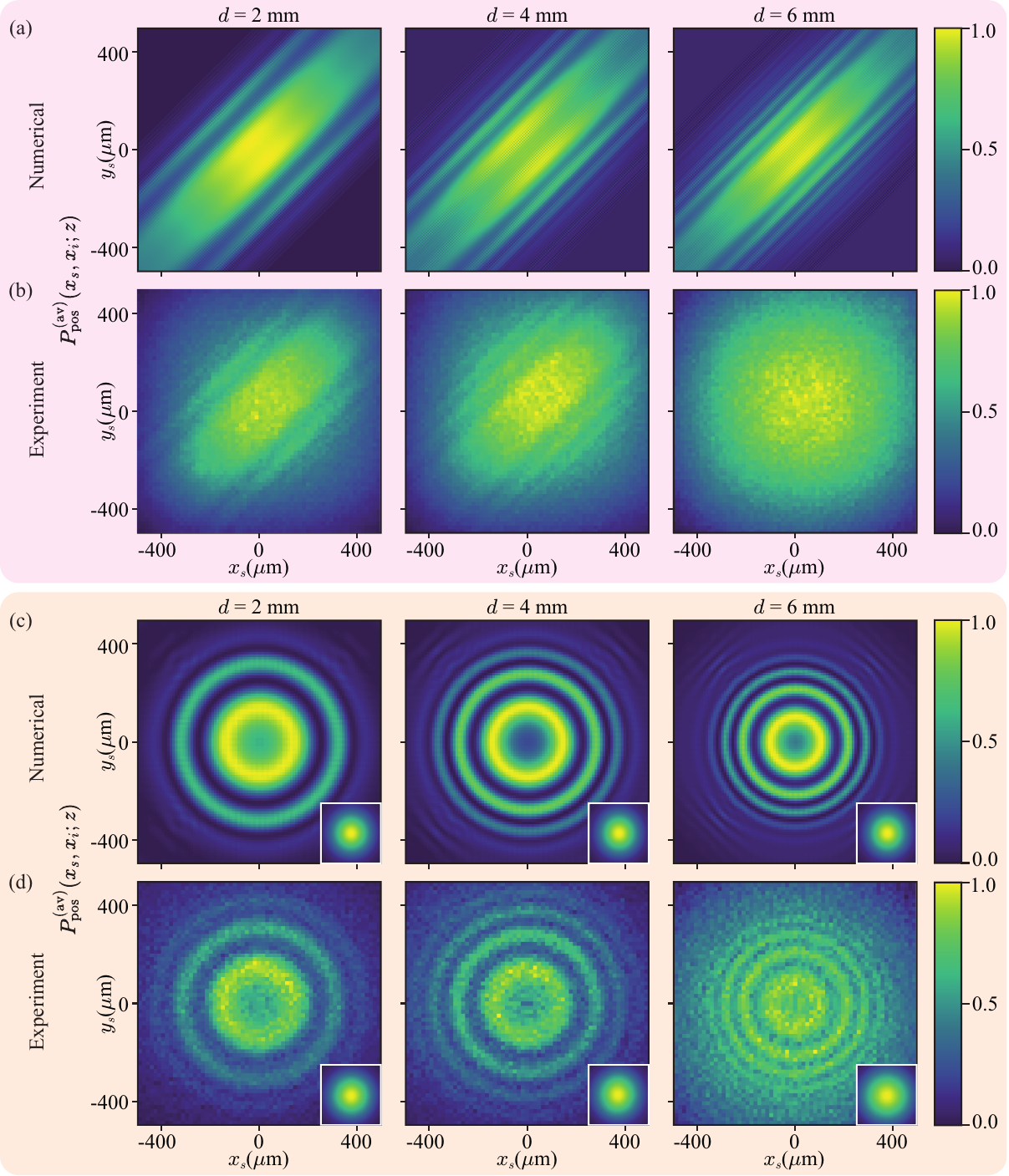}
\caption{(a), (b) Numerically evaluated and experimental measured plots of $P_{\rm pos}^{(\rm av)}(x_s, x_i; z)$ at various $d$ with $\theta_p=92.9^{\circ}$ and $z=7.5$ mm from the center of the two crystals. (c), (d) Numerically evaluated and experimental measured plots of $P_{\rm pos}(x_s, y_s|0, 0; z)$ at various $d$ with $\theta_p=92.9^{\circ}$ and $z=7.5$ mm from the center of the two crystals.}
\label{fig:2 crystal}
\end{figure*}
The position-momentum entanglement of the SPDC photons can be certified either through the violation of the conditional Heisenberg uncertainty relation \cite{bhattacharjee2022sciadv, howell2004prl} or by demonstrating a positive value for the minimum of entanglement of formation$E_f^{(\rm min)}$ \cite{prevedel2011natphy,schneeloch2019natcomm}. For violating the conditional Heisenberg uncertainty relation, one needs to calculate the conditional uncertainties, which can be defined when the two-photon probability distribution functions are singly-peaked and have well-defined widths. However, in our present work, the concerned functions have multiple peaks, which makes it difficult to define the width. Therefore, for certifying entanglement in our work, we use the minimum of entanglement of formation  $E_f^{(\rm min)}$ \cite{prevedel2011natphy,schneeloch2019natcomm}, which is given by \cite{berta2010natphy}
\begin{equation}\label{eq:ef}
E_{f}^{(\rm min)} = 2\log_{2}(M) - H(\bm{X}_s|\bm{X}_i) - H(\bm{K}_s|\bm{K}_i).
\end{equation}
Here $M$ is the dimensionality of the Hilbert space, and ($\boldsymbol{X}$) and ($\boldsymbol{K}$) are the discrete position and transverse-momentum basis vectors corresponding to the continuous basis vectors $\bm\rho$ and $\bm q$, respectively \cite{schneeloch2018pra}. $H(\bm{X}_s|\bm{X}_i)$ is the conditional von Neumann entropies in the position basis  and can be expressed as:
\begin{align*}
H\left(\boldsymbol{X}_s|\boldsymbol{X_i}\right)=H\left(\boldsymbol{X}_s, \boldsymbol{X}_i\right)-H\left(\boldsymbol{X}_i\right),
\end{align*} 
where
\begin{align*}
&H\left(\boldsymbol{X}_s, \boldsymbol{X}_i\right)
=-\sum_{\boldsymbol{X}_s} \sum_{\boldsymbol{X}_s} P_{\rm pos}\left(\boldsymbol{X}_s, \boldsymbol{X}_i; z\right) \ln _2 P_{\rm pos}\left(\boldsymbol{X}_s, \boldsymbol{X}_i; z\right) \\
&H\left(\boldsymbol{X}_i\right)
=-\sum_{\boldsymbol{X}_i} (\sum_{\boldsymbol{X}_s} P_{\rm pos}\left(\boldsymbol{X}_s, \boldsymbol{X}_i; z\right) ) \\
& \qquad\qquad\qquad\qquad\qquad \times\ln _2 (\sum_{\boldsymbol{X}_s} P_{\rm pos}\left(\boldsymbol{X}_s, \boldsymbol{X}_i; z\right)).
\end{align*}

Here, $P_{\rm pos}\left(\boldsymbol{X}_s, \boldsymbol{X}_i; z\right))$ is the discrete probability distribution function corresponding to the continuous two-photon probability distribution function $P_{\rm pos}(\boldsymbol{\rho}_s, \boldsymbol{\rho}_i; z)$. We note that, in an experiment situation, $P_{\rm pos}\left(\boldsymbol{X}_s, \boldsymbol{X}_i; z\right))$ is measured using a multi-pixel camera. The camera pixels provide the discrete space, and the value of $P_{\rm pos}(\boldsymbol{\rho}_s, \boldsymbol{\rho}_i; z)$ at a pair of pixels located at $\boldsymbol{X}_s$ and $\boldsymbol{X}_i$, averaged over the pixel area, is taken as $P_{\rm pos}\left(\boldsymbol{X}_s, \boldsymbol{X}_i; z\right))$. In a similar manner, we can write the conditional von Neumann entropy $H(\bm{K}_s|\bm{K}_i)$ in the momentum basis and express that in terms of the discrete probability distribution function $P_{\rm mom}\left(\boldsymbol{K}_s, \boldsymbol{K}_i; z\right))$. A positive value of $E_f^{(\rm min)}$ implies entanglement, and we use it as the criterion for verifying entanglement \cite{prevedel2011natphy,schneeloch2019natcomm}.

\section{Results}

\subsection{Experimental Entanglement verification}

In our experiments, we measure $E_f^{(\rm min)}$ for verifying entanglement. First of all, in order to find the $z$ and $\theta_p$ ranges over which entanglement remains present, we numerically evaluate $E_f^{(\rm min)}$ using Eq.~(\ref{eq:ef}) as a function of $z$ and $\theta_p$. Figure \ref{min-EOF} shows the plot of $E_f^{(\rm min)}$ as a function of $z$ for $\theta_p=32.9^{\circ}$. We find that $E_f^{(\rm min)}$ remains positive upto $z=20$ mm, which implies the presence of entanglement upto $z=20$ mm. We also simulate $E_f^{(\rm min)}$ at $z=5$ mm for a range of $\theta_p$ values between $32.9^{\circ}$ and $33.0^{\circ}$, and we find that within this range $E_f^{(\rm min)}$ remains positive. Therefore, we perform our experiments at $z=5$ mm and with $\theta_p$ values ranging between $32.9^{\circ}$ and $33.0^{\circ}$ to ensure that the generated two-photon field remains entangled in position-momentum degree of freedom.

\subsection{Observation of structured correlation}	
	
Figure \ref{fig:exp set up} shows the experimental setup. A continuous wave UV laser of wavelength 355 nm, power 5 mW, and beam waist 507 $\mu$m is used for producing Type-I SPDC photons using $\beta$-Barium Borate (BBO) crystals. The down-converted photons are detected using an Andor iXon Ultra-897 EMCCD camera. A 10 nm band pass filter centered at 710 nm is used before the camera. The camera operates at an EM gain of 1000, a temperature of $-60^{\circ}$C, and a frame rate of ~$170$ Hz. It is used for detecting the SPDC photons in coincidence by following the procedure detailed in Refs.~\cite{edgar2012natcomm,defienne2018prl,reichert2018scirep, ndagano2020npjqinfo}. For measuring position correlations, at different $z$-planes and $\theta_p$ values, we use a $4f$ imaging system with a lens of focal length 10 cm and with the EMCCD camera placed at the image plane, as shown in Fig \ref{fig:exp set up}(a). For measuring the transverse momentum correlations, we use a $2f$ imaging system in a manner that the camera gets positioned at the Fourier plane of the desired $z$-plane, as shown in Fig \ref{fig:exp set up}(b). 

We employ two different methods for structuring the correlation function.  In the first method, SPDC is generated using a single BBO crystal of length $L=5$ mm and the structuring is achieved by changing $\theta_p$. The phase-matching function in this case is given by Eq.~(\ref{eq:single crystal phi}), and Figure \ref{fig:1 crystal} presents the numerically simulated and experimentally measured probabilities $P_{\rm pos}^{(\rm av)}(x_s, x_i; z)$ and $P_{\rm pos}(x_s, y_s|0, 0; z)$. We find that at the collinear phase matching condition, that is, at $\theta_p=32.9^{\circ}$, we get the familiar result of $P_{\rm pos}^{(\rm av)}(x_s, x_i; z)$ showing anti-diagonal correlations with $P_{\rm pos}(x_s, y_s|0, 0; z)$ being in the form of a two-dimensional function centered at the origin \cite{howell2004prl, bhattacharjee2022sciadv}. However, as $\theta_p$ increases and the phase matching becomes more noncollinear, the diagonal form of the $P_{\rm pos}^{(\rm av)}(x_s, x_i; z)$ bifurcates, and as a result, $P_{\rm pos}(x_s, y_s|0, 0; z)$ takes the shape of an annulus. We note that within the range of $0-20$ mm over which the $E_f^{(\rm min)}$ remains positive, as illustrated in Fig.~\ref{min-EOF}(a), the dependence of $P_{\rm pos}^{(\rm av)}(x_s, x_i; z)$ on $\theta_p$ at different $z$ qualitatively remain the same, except for the overall widening of the function. We further note that the structuring is seen only in the two-photon probability distribution functions but not in the one-photon probability distribution function, as presented as the insets in Figure~\ref{fig:1 crystal}(b). Next, we measure the transverse momentum correlations and presents the numerically simulated and experimentally measured $P_{\rm mom}^{(\rm av)}(q_{xs}, q_{xi}; z)$ at different $\theta_p$ values in Fig.~\ref{fig:mom corr}. Using the results reported in Figs.~\ref{fig:1 crystal} and \ref{fig:mom corr}, we calculate $E_f^{(\rm min)}$ and present it in TABLE \ref{table:singleq}. The positive values of $E_f^{(\rm min)}$ observed in our experiment confirms the presence of position-momentum entanglement in the generated two-photon state, and overall, we find very good match between simulations and experiments. 
 \begin{table}[t]
		\begin{minipage}[t]{0.4\textwidth}
			\centering
			\begin{tabular}{ |c|c|c| }
				\hline
				$\theta_p$ & $E_f$ (theory) & $E_f$ (experiment) \\
				\hline
				$32.90^{\circ}$ & 1.89 & 1.03 \\
				\hline
				$32.92^{\circ}$ & 2.67 & 1.11 \\
				\hline
				$32.94^{\circ}$ & 2.06 & 1.02 \\
				\hline
				$32.96^{\circ}$ & 1.79 & 1.00 \\
				\hline
				$32.98^{\circ}$ & 1.49 & 1.01 \\
				\hline
			\end{tabular}
			\caption{Theoretical and experimentally obtained values for $E_f$ for the single crystal configuration. ($z=5$ mm)}
			\label{table:singleq}
		\end{minipage}
		\hspace{0.5cm}
		\begin{minipage}[t]{0.4\textwidth}
			\centering
			\begin{tabular}{ |c|c|c| }
				\hline
				$d$ (mm) & $E_f$ (theory) & $E_f$ (experiment) \\
				\hline
				$2$ & 2.18 & 1.14 \\
				\hline
				$4$ & 1.97 & 1.03 \\
				\hline
				$6$ & 1.77 & 1.00 \\
				\hline
			\end{tabular}
			\caption{Theoretical and experimentally obtained values for $E_f$ for the double crystal configuration. ($z=7.5$ mm and $\theta_p=32.93^{\circ}$)}
			\label{table:doubleq}
		\end{minipage}
	\end{table}

In the second method of structuring the correlation functions, we generate SPDC with a pair of BBO crystals of thickness $L=1$ mm each, separated by $d=2$, $4$ and $6$ mm.  The phase-matching angle is kept at $\theta_p=32.93^{\circ}$ for both the BBO crystals with $z = 7.5$ mm, measured from the middle of the two crystals. The corresponding phase-matching function is given by Eq.~(\ref{eq:double crystal phi}). Figure \ref{fig:2 crystal} presents the numerically simulated and experimentally measured probabilities $P_{\rm pos}^{(\rm av)}(x_s, x_i; z)$ and $P_{\rm pos}(x_s, y_s|0, 0; z)$. We find that this method generates more complex structures compared to the first method. The reason for this is that the two-photon field produced from the first crystal interferes with that from the second crystal and produces interference. The distance $d$ between the two crystals determines the richness of the crystal. We note that just like in the first method, all the structuring is seen only in the two-photon probability functions but not in the one-photon detection probability, presented as the insets in Figure~\ref{fig:2 crystal}(b).  This is because of the fact that although the two-photon field generated from the two crystals are mutually temporally coherent with the two-photon coherence length in the range of a few mm, the single-photon fields from the two crystals are mutually temporally incoherent with the temporal coherence lengths of the field being of the order of only 100 microns \cite{jha2008pra, kulkarni2017josab}. The numerically calculated and experimentally measured values of $E_f^{(\rm min}$ are presented in TABLE \ref{table:doubleq}. We note that, although the positive values of $E_f^{(\rm min}$  confirms the presence of position-momentum entanglement, the match between the theory and experiment is not as good as in Fig.~\ref{fig:2 crystal}, especially at $d=6$ mm. This is primarily because of the richness in the structure, which due to its inherent noise, the EMCCD-based coincidence measurement is not able to resolve. A multi-pixel single-photon camera with lower inherent noise should be able to measure these structures with better resolution. Furthermore, we note that, in our experiments the pump power used is only about 5 mW. At this field strength, the possibility of generating higher-photon states is negligible \cite{schneeloch2019jopt}, and therefore what we observe are primarily two-photon effects.

	\section{Conclusion}

In this article, we have reported experimental generation of two-photon fields with various structures in their spatial correlations. We have demonstrated that such fields can be produced in the near-field region of the SPDC crystal by manipulating the phase-matching condition through either changing the phase-matching angle or having multiple SPDC crystals. We have confirmed the entanglement of our structured fields through entanglement of formation. This work can have important implications for quantum imaging and microscopy.

	\begin{acknowledgments}
		We acknowledge financial support from the Science and Engineering Research Board, Government of India, through grants STR/2021/000035 and CRG/2022/003070, and from the Department of Science and Technology, Government of India, through Grant DST/ICPS/QuST/Theme-1/2019).  R.P. thanks the Prime Minister's Research Fellowship, Government of India and S.K. thanks the University Grant Commission, Government of India, for financial support.

	\end{acknowledgments}


	\bibliography{mybibiliographyref}

\begin{thebibliography}{62}%
\makeatletter
\providecommand \@ifxundefined [1]{%
 \@ifx{#1\undefined}
}%
\providecommand \@ifnum [1]{%
 \ifnum #1\expandafter \@firstoftwo
 \else \expandafter \@secondoftwo
 \fi
}%
\providecommand \@ifx [1]{%
 \ifx #1\expandafter \@firstoftwo
 \else \expandafter \@secondoftwo
 \fi
}%
\providecommand \natexlab [1]{#1}%
\providecommand \enquote  [1]{``#1''}%
\providecommand \bibnamefont  [1]{#1}%
\providecommand \bibfnamefont [1]{#1}%
\providecommand \citenamefont [1]{#1}%
\providecommand \href@noop [0]{\@secondoftwo}%
\providecommand \href [0]{\begingroup \@sanitize@url \@href}%
\providecommand \@href[1]{\@@startlink{#1}\@@href}%
\providecommand \@@href[1]{\endgroup#1\@@endlink}%
\providecommand \@sanitize@url [0]{\catcode `\\12\catcode `\$12\catcode
  `\&12\catcode `\#12\catcode `\^12\catcode `\_12\catcode `\%12\relax}%
\providecommand \@@startlink[1]{}%
\providecommand \@@endlink[0]{}%
\providecommand \url  [0]{\begingroup\@sanitize@url \@url }%
\providecommand \@url [1]{\endgroup\@href {#1}{\urlprefix }}%
\providecommand \urlprefix  [0]{URL }%
\providecommand \Eprint [0]{\href }%
\providecommand \doibase [0]{https://doi.org/}%
\providecommand \selectlanguage [0]{\@gobble}%
\providecommand \bibinfo  [0]{\@secondoftwo}%
\providecommand \bibfield  [0]{\@secondoftwo}%
\providecommand \translation [1]{[#1]}%
\providecommand \BibitemOpen [0]{}%
\providecommand \bibitemStop [0]{}%
\providecommand \bibitemNoStop [0]{.\EOS\space}%
\providecommand \EOS [0]{\spacefactor3000\relax}%
\providecommand \BibitemShut  [1]{\csname bibitem#1\endcsname}%
\let\auto@bib@innerbib\@empty
\bibitem [{\citenamefont {Hell}\ and\ \citenamefont
  {Wichmann}(1994)}]{hell1994optlett}%
  \BibitemOpen
  \bibfield  {author} {\bibinfo {author} {\bibfnamefont {S.~W.}\ \bibnamefont
  {Hell}}\ and\ \bibinfo {author} {\bibfnamefont {J.}~\bibnamefont
  {Wichmann}},\ }\bibfield  {title} {\bibinfo {title} {Breaking the diffraction
  resolution limit by stimulated emission: stimulated-emission-depletion
  fluorescence microscopy},\ }\href@noop {} {\bibfield  {journal} {\bibinfo
  {journal} {Optics letters}\ }\textbf {\bibinfo {volume} {19}},\ \bibinfo
  {pages} {780} (\bibinfo {year} {1994})}\BibitemShut {NoStop}%
\bibitem [{\citenamefont {Samanta}\ and\ \citenamefont
  {Joseph}(2021)}]{samanta2021jopt}%
  \BibitemOpen
  \bibfield  {author} {\bibinfo {author} {\bibfnamefont {K.}~\bibnamefont
  {Samanta}}\ and\ \bibinfo {author} {\bibfnamefont {J.}~\bibnamefont
  {Joseph}},\ }\bibfield  {title} {\bibinfo {title} {An overview of structured
  illumination microscopy: recent advances and perspectives},\ }\href@noop {}
  {\bibfield  {journal} {\bibinfo  {journal} {Journal of Optics}\ }\textbf
  {\bibinfo {volume} {23}},\ \bibinfo {pages} {123002} (\bibinfo {year}
  {2021})}\BibitemShut {NoStop}%
\bibitem [{\citenamefont {Curcio}\ \emph {et~al.}(2020)\citenamefont {Curcio},
  \citenamefont {Alem{\'a}n-Casta{\~n}eda}, \citenamefont {Brown},
  \citenamefont {Brasselet},\ and\ \citenamefont {Alonso}}]{curcio2020natcomm}%
  \BibitemOpen
  \bibfield  {author} {\bibinfo {author} {\bibfnamefont {V.}~\bibnamefont
  {Curcio}}, \bibinfo {author} {\bibfnamefont {L.~A.}\ \bibnamefont
  {Alem{\'a}n-Casta{\~n}eda}}, \bibinfo {author} {\bibfnamefont {T.~G.}\
  \bibnamefont {Brown}}, \bibinfo {author} {\bibfnamefont {S.}~\bibnamefont
  {Brasselet}},\ and\ \bibinfo {author} {\bibfnamefont {M.~A.}\ \bibnamefont
  {Alonso}},\ }\bibfield  {title} {\bibinfo {title} {Birefringent fourier
  filtering for single molecule coordinate and height super-resolution imaging
  with dithering and orientation},\ }\href@noop {} {\bibfield  {journal}
  {\bibinfo  {journal} {Nature communications}\ }\textbf {\bibinfo {volume}
  {11}},\ \bibinfo {pages} {5307} (\bibinfo {year} {2020})}\BibitemShut
  {NoStop}%
\bibitem [{\citenamefont {Rosales-Guzm{\'a}n}\ \emph
  {et~al.}(2013)\citenamefont {Rosales-Guzm{\'a}n}, \citenamefont {Hermosa},
  \citenamefont {Belmonte},\ and\ \citenamefont {Torres}}]{rosales2013scirep}%
  \BibitemOpen
  \bibfield  {author} {\bibinfo {author} {\bibfnamefont {C.}~\bibnamefont
  {Rosales-Guzm{\'a}n}}, \bibinfo {author} {\bibfnamefont {N.}~\bibnamefont
  {Hermosa}}, \bibinfo {author} {\bibfnamefont {A.}~\bibnamefont {Belmonte}},\
  and\ \bibinfo {author} {\bibfnamefont {J.~P.}\ \bibnamefont {Torres}},\
  }\bibfield  {title} {\bibinfo {title} {Experimental detection of transverse
  particle movement with structured light},\ }\href@noop {} {\bibfield
  {journal} {\bibinfo  {journal} {Scientific reports}\ }\textbf {\bibinfo
  {volume} {3}},\ \bibinfo {pages} {2815} (\bibinfo {year} {2013})}\BibitemShut
  {NoStop}%
\bibitem [{\citenamefont {Liang}\ \emph {et~al.}(2017)\citenamefont {Liang},
  \citenamefont {Wu}, \citenamefont {Wang}, \citenamefont {Li}, \citenamefont
  {Cai},\ and\ \citenamefont {Ponomarenko}}]{liang2017optexp}%
  \BibitemOpen
  \bibfield  {author} {\bibinfo {author} {\bibfnamefont {C.}~\bibnamefont
  {Liang}}, \bibinfo {author} {\bibfnamefont {G.}~\bibnamefont {Wu}}, \bibinfo
  {author} {\bibfnamefont {F.}~\bibnamefont {Wang}}, \bibinfo {author}
  {\bibfnamefont {W.}~\bibnamefont {Li}}, \bibinfo {author} {\bibfnamefont
  {Y.}~\bibnamefont {Cai}},\ and\ \bibinfo {author} {\bibfnamefont {S.~A.}\
  \bibnamefont {Ponomarenko}},\ }\bibfield  {title} {\bibinfo {title}
  {Overcoming the classical rayleigh diffraction limit by controlling two-point
  correlations of partially coherent light sources},\ }\href@noop {} {\bibfield
   {journal} {\bibinfo  {journal} {Optics Express}\ }\textbf {\bibinfo {volume}
  {25}},\ \bibinfo {pages} {28352} (\bibinfo {year} {2017})}\BibitemShut
  {NoStop}%
\bibitem [{\citenamefont {Shen}\ \emph {et~al.}(2022)\citenamefont {Shen},
  \citenamefont {Pidishety}, \citenamefont {Nape},\ and\ \citenamefont
  {Dudley}}]{shen2022jopt}%
  \BibitemOpen
  \bibfield  {author} {\bibinfo {author} {\bibfnamefont {Y.}~\bibnamefont
  {Shen}}, \bibinfo {author} {\bibfnamefont {S.}~\bibnamefont {Pidishety}},
  \bibinfo {author} {\bibfnamefont {I.}~\bibnamefont {Nape}},\ and\ \bibinfo
  {author} {\bibfnamefont {A.}~\bibnamefont {Dudley}},\ }\bibfield  {title}
  {\bibinfo {title} {Self-healing of structured light: a review},\ }\href@noop
  {} {\bibfield  {journal} {\bibinfo  {journal} {Journal of Optics}\ }\textbf
  {\bibinfo {volume} {24}},\ \bibinfo {pages} {103001} (\bibinfo {year}
  {2022})}\BibitemShut {NoStop}%
\bibitem [{\citenamefont {Bozinovic}\ \emph {et~al.}(2013)\citenamefont
  {Bozinovic}, \citenamefont {Yue}, \citenamefont {Ren}, \citenamefont {Tur},
  \citenamefont {Kristensen}, \citenamefont {Huang}, \citenamefont {Willner},\
  and\ \citenamefont {Ramachandran}}]{bozinovic2013science}%
  \BibitemOpen
  \bibfield  {author} {\bibinfo {author} {\bibfnamefont {N.}~\bibnamefont
  {Bozinovic}}, \bibinfo {author} {\bibfnamefont {Y.}~\bibnamefont {Yue}},
  \bibinfo {author} {\bibfnamefont {Y.}~\bibnamefont {Ren}}, \bibinfo {author}
  {\bibfnamefont {M.}~\bibnamefont {Tur}}, \bibinfo {author} {\bibfnamefont
  {P.}~\bibnamefont {Kristensen}}, \bibinfo {author} {\bibfnamefont
  {H.}~\bibnamefont {Huang}}, \bibinfo {author} {\bibfnamefont {A.~E.}\
  \bibnamefont {Willner}},\ and\ \bibinfo {author} {\bibfnamefont
  {S.}~\bibnamefont {Ramachandran}},\ }\bibfield  {title} {\bibinfo {title}
  {Terabit-scale orbital angular momentum mode division multiplexing in
  fibers},\ }\href {https://doi.org/10.1126/science.1237861} {\bibfield
  {journal} {\bibinfo  {journal} {Science}\ }\textbf {\bibinfo {volume}
  {340}},\ \bibinfo {pages} {1545} (\bibinfo {year} {2013})}\BibitemShut
  {NoStop}%
\bibitem [{\citenamefont {Wang}\ \emph {et~al.}(2012)\citenamefont {Wang},
  \citenamefont {Yang}, \citenamefont {Fazal}, \citenamefont {Ahmed},
  \citenamefont {Yan}, \citenamefont {Huang}, \citenamefont {Ren},
  \citenamefont {Yue}, \citenamefont {Dolinar}, \citenamefont {Tur} \emph
  {et~al.}}]{wang2012science}%
  \BibitemOpen
  \bibfield  {author} {\bibinfo {author} {\bibfnamefont {J.}~\bibnamefont
  {Wang}}, \bibinfo {author} {\bibfnamefont {J.-Y.}\ \bibnamefont {Yang}},
  \bibinfo {author} {\bibfnamefont {I.~M.}\ \bibnamefont {Fazal}}, \bibinfo
  {author} {\bibfnamefont {N.}~\bibnamefont {Ahmed}}, \bibinfo {author}
  {\bibfnamefont {Y.}~\bibnamefont {Yan}}, \bibinfo {author} {\bibfnamefont
  {H.}~\bibnamefont {Huang}}, \bibinfo {author} {\bibfnamefont
  {Y.}~\bibnamefont {Ren}}, \bibinfo {author} {\bibfnamefont {Y.}~\bibnamefont
  {Yue}}, \bibinfo {author} {\bibfnamefont {S.}~\bibnamefont {Dolinar}},
  \bibinfo {author} {\bibfnamefont {M.}~\bibnamefont {Tur}}, \emph {et~al.},\
  }\bibfield  {title} {\bibinfo {title} {Terabit free-space data transmission
  employing orbital angular momentum multiplexing},\ }\href@noop {} {\bibfield
  {journal} {\bibinfo  {journal} {Nature photonics}\ }\textbf {\bibinfo
  {volume} {6}},\ \bibinfo {pages} {488} (\bibinfo {year} {2012})}\BibitemShut
  {NoStop}%
\bibitem [{\citenamefont {Forbes}\ \emph {et~al.}(2021)\citenamefont {Forbes},
  \citenamefont {de~Oliveira},\ and\ \citenamefont
  {Dennis}}]{forbes2021natphot}%
  \BibitemOpen
  \bibfield  {author} {\bibinfo {author} {\bibfnamefont {A.}~\bibnamefont
  {Forbes}}, \bibinfo {author} {\bibfnamefont {M.}~\bibnamefont
  {de~Oliveira}},\ and\ \bibinfo {author} {\bibfnamefont {M.~R.}\ \bibnamefont
  {Dennis}},\ }\bibfield  {title} {\bibinfo {title} {Structured light},\
  }\href@noop {} {\bibfield  {journal} {\bibinfo  {journal} {Nature Photonics}\
  }\textbf {\bibinfo {volume} {15}},\ \bibinfo {pages} {253} (\bibinfo {year}
  {2021})}\BibitemShut {NoStop}%
\bibitem [{\citenamefont {Rubinsztein-Dunlop}\ \emph
  {et~al.}(2016)\citenamefont {Rubinsztein-Dunlop}, \citenamefont {Forbes},
  \citenamefont {Berry}, \citenamefont {Dennis}, \citenamefont {Andrews},
  \citenamefont {Mansuripur}, \citenamefont {Denz}, \citenamefont {Alpmann},
  \citenamefont {Banzer}, \citenamefont {Bauer} \emph
  {et~al.}}]{rubinsztein2016jopt}%
  \BibitemOpen
  \bibfield  {author} {\bibinfo {author} {\bibfnamefont {H.}~\bibnamefont
  {Rubinsztein-Dunlop}}, \bibinfo {author} {\bibfnamefont {A.}~\bibnamefont
  {Forbes}}, \bibinfo {author} {\bibfnamefont {M.~V.}\ \bibnamefont {Berry}},
  \bibinfo {author} {\bibfnamefont {M.~R.}\ \bibnamefont {Dennis}}, \bibinfo
  {author} {\bibfnamefont {D.~L.}\ \bibnamefont {Andrews}}, \bibinfo {author}
  {\bibfnamefont {M.}~\bibnamefont {Mansuripur}}, \bibinfo {author}
  {\bibfnamefont {C.}~\bibnamefont {Denz}}, \bibinfo {author} {\bibfnamefont
  {C.}~\bibnamefont {Alpmann}}, \bibinfo {author} {\bibfnamefont
  {P.}~\bibnamefont {Banzer}}, \bibinfo {author} {\bibfnamefont
  {T.}~\bibnamefont {Bauer}}, \emph {et~al.},\ }\bibfield  {title} {\bibinfo
  {title} {Roadmap on structured light},\ }\href@noop {} {\bibfield  {journal}
  {\bibinfo  {journal} {Journal of Optics}\ }\textbf {\bibinfo {volume} {19}},\
  \bibinfo {pages} {013001} (\bibinfo {year} {2016})}\BibitemShut {NoStop}%
\bibitem [{\citenamefont {He}\ \emph {et~al.}(2022)\citenamefont {He},
  \citenamefont {Shen},\ and\ \citenamefont {Forbes}}]{he2022lsciapp}%
  \BibitemOpen
  \bibfield  {author} {\bibinfo {author} {\bibfnamefont {C.}~\bibnamefont
  {He}}, \bibinfo {author} {\bibfnamefont {Y.}~\bibnamefont {Shen}},\ and\
  \bibinfo {author} {\bibfnamefont {A.}~\bibnamefont {Forbes}},\ }\bibfield
  {title} {\bibinfo {title} {Towards higher-dimensional structured light},\
  }\href@noop {} {\bibfield  {journal} {\bibinfo  {journal} {Light: Science \&
  Applications}\ }\textbf {\bibinfo {volume} {11}},\ \bibinfo {pages} {205}
  (\bibinfo {year} {2022})}\BibitemShut {NoStop}%
\bibitem [{\citenamefont {D'ambrosio}\ \emph {et~al.}(2012)\citenamefont
  {D'ambrosio}, \citenamefont {Nagali}, \citenamefont {Walborn}, \citenamefont
  {Aolita}, \citenamefont {Slussarenko}, \citenamefont {Marrucci},\ and\
  \citenamefont {Sciarrino}}]{dambrosio2012natcomm}%
  \BibitemOpen
  \bibfield  {author} {\bibinfo {author} {\bibfnamefont {V.}~\bibnamefont
  {D'ambrosio}}, \bibinfo {author} {\bibfnamefont {E.}~\bibnamefont {Nagali}},
  \bibinfo {author} {\bibfnamefont {S.~P.}\ \bibnamefont {Walborn}}, \bibinfo
  {author} {\bibfnamefont {L.}~\bibnamefont {Aolita}}, \bibinfo {author}
  {\bibfnamefont {S.}~\bibnamefont {Slussarenko}}, \bibinfo {author}
  {\bibfnamefont {L.}~\bibnamefont {Marrucci}},\ and\ \bibinfo {author}
  {\bibfnamefont {F.}~\bibnamefont {Sciarrino}},\ }\bibfield  {title} {\bibinfo
  {title} {Complete experimental toolbox for alignment-free quantum
  communication},\ }\href@noop {} {\bibfield  {journal} {\bibinfo  {journal}
  {Nature communications}\ }\textbf {\bibinfo {volume} {3}},\ \bibinfo {pages}
  {961} (\bibinfo {year} {2012})}\BibitemShut {NoStop}%
\bibitem [{\citenamefont {Marcikic}\ \emph {et~al.}(2003)\citenamefont
  {Marcikic}, \citenamefont {De~Riedmatten}, \citenamefont {Tittel},
  \citenamefont {Zbinden},\ and\ \citenamefont {Gisin}}]{marcikic2003nature}%
  \BibitemOpen
  \bibfield  {author} {\bibinfo {author} {\bibfnamefont {I.}~\bibnamefont
  {Marcikic}}, \bibinfo {author} {\bibfnamefont {H.}~\bibnamefont
  {De~Riedmatten}}, \bibinfo {author} {\bibfnamefont {W.}~\bibnamefont
  {Tittel}}, \bibinfo {author} {\bibfnamefont {H.}~\bibnamefont {Zbinden}},\
  and\ \bibinfo {author} {\bibfnamefont {N.}~\bibnamefont {Gisin}},\ }\bibfield
   {title} {\bibinfo {title} {Long-distance teleportation of qubits at
  telecommunication wavelengths},\ }\href@noop {} {\bibfield  {journal}
  {\bibinfo  {journal} {Nature}\ }\textbf {\bibinfo {volume} {421}},\ \bibinfo
  {pages} {509} (\bibinfo {year} {2003})}\BibitemShut {NoStop}%
\bibitem [{\citenamefont {Ecker}\ \emph {et~al.}(2019)\citenamefont {Ecker},
  \citenamefont {Bouchard}, \citenamefont {Bulla}, \citenamefont {Brandt},
  \citenamefont {Kohout}, \citenamefont {Steinlechner}, \citenamefont
  {Fickler}, \citenamefont {Malik}, \citenamefont {Guryanova}, \citenamefont
  {Ursin} \emph {et~al.}}]{ecker2019prx}%
  \BibitemOpen
  \bibfield  {author} {\bibinfo {author} {\bibfnamefont {S.}~\bibnamefont
  {Ecker}}, \bibinfo {author} {\bibfnamefont {F.}~\bibnamefont {Bouchard}},
  \bibinfo {author} {\bibfnamefont {L.}~\bibnamefont {Bulla}}, \bibinfo
  {author} {\bibfnamefont {F.}~\bibnamefont {Brandt}}, \bibinfo {author}
  {\bibfnamefont {O.}~\bibnamefont {Kohout}}, \bibinfo {author} {\bibfnamefont
  {F.}~\bibnamefont {Steinlechner}}, \bibinfo {author} {\bibfnamefont
  {R.}~\bibnamefont {Fickler}}, \bibinfo {author} {\bibfnamefont
  {M.}~\bibnamefont {Malik}}, \bibinfo {author} {\bibfnamefont
  {Y.}~\bibnamefont {Guryanova}}, \bibinfo {author} {\bibfnamefont
  {R.}~\bibnamefont {Ursin}}, \emph {et~al.},\ }\bibfield  {title} {\bibinfo
  {title} {Overcoming noise in entanglement distribution},\ }\href@noop {}
  {\bibfield  {journal} {\bibinfo  {journal} {Physical Review X}\ }\textbf
  {\bibinfo {volume} {9}},\ \bibinfo {pages} {041042} (\bibinfo {year}
  {2019})}\BibitemShut {NoStop}%
\bibitem [{\citenamefont {Ekert}(1991)}]{ekert1991prl}%
  \BibitemOpen
  \bibfield  {author} {\bibinfo {author} {\bibfnamefont {A.~K.}\ \bibnamefont
  {Ekert}},\ }\bibfield  {title} {\bibinfo {title} {Quantum cryptography based
  on {B}ell\char39{}s theorem},\ }\href
  {https://doi.org/10.1103/PhysRevLett.67.661} {\bibfield  {journal} {\bibinfo
  {journal} {Phys. Rev. Lett.}\ }\textbf {\bibinfo {volume} {67}},\ \bibinfo
  {pages} {661} (\bibinfo {year} {1991})}\BibitemShut {NoStop}%
\bibitem [{\citenamefont {Jennewein}\ \emph {et~al.}(2000)\citenamefont
  {Jennewein}, \citenamefont {Simon}, \citenamefont {Weihs}, \citenamefont
  {Weinfurter},\ and\ \citenamefont {Zeilinger}}]{jennewein2000prl}%
  \BibitemOpen
  \bibfield  {author} {\bibinfo {author} {\bibfnamefont {T.}~\bibnamefont
  {Jennewein}}, \bibinfo {author} {\bibfnamefont {C.}~\bibnamefont {Simon}},
  \bibinfo {author} {\bibfnamefont {G.}~\bibnamefont {Weihs}}, \bibinfo
  {author} {\bibfnamefont {H.}~\bibnamefont {Weinfurter}},\ and\ \bibinfo
  {author} {\bibfnamefont {A.}~\bibnamefont {Zeilinger}},\ }\bibfield  {title}
  {\bibinfo {title} {Quantum cryptography with entangled photons},\ }\href
  {https://doi.org/10.1103/PhysRevLett.84.4729} {\bibfield  {journal} {\bibinfo
   {journal} {Phys. Rev. Lett.}\ }\textbf {\bibinfo {volume} {84}},\ \bibinfo
  {pages} {4729} (\bibinfo {year} {2000})}\BibitemShut {NoStop}%
\bibitem [{\citenamefont {Liao}\ \emph {et~al.}(2017)\citenamefont {Liao},
  \citenamefont {Yong}, \citenamefont {Liu}, \citenamefont {Shentu},
  \citenamefont {Li}, \citenamefont {Lin}, \citenamefont {Dai}, \citenamefont
  {Zhao}, \citenamefont {Li}, \citenamefont {Guan} \emph
  {et~al.}}]{liao2017natcomm}%
  \BibitemOpen
  \bibfield  {author} {\bibinfo {author} {\bibfnamefont {S.-K.}\ \bibnamefont
  {Liao}}, \bibinfo {author} {\bibfnamefont {H.-L.}\ \bibnamefont {Yong}},
  \bibinfo {author} {\bibfnamefont {C.}~\bibnamefont {Liu}}, \bibinfo {author}
  {\bibfnamefont {G.-L.}\ \bibnamefont {Shentu}}, \bibinfo {author}
  {\bibfnamefont {D.-D.}\ \bibnamefont {Li}}, \bibinfo {author} {\bibfnamefont
  {J.}~\bibnamefont {Lin}}, \bibinfo {author} {\bibfnamefont {H.}~\bibnamefont
  {Dai}}, \bibinfo {author} {\bibfnamefont {S.-Q.}\ \bibnamefont {Zhao}},
  \bibinfo {author} {\bibfnamefont {B.}~\bibnamefont {Li}}, \bibinfo {author}
  {\bibfnamefont {J.-Y.}\ \bibnamefont {Guan}}, \emph {et~al.},\ }\bibfield
  {title} {\bibinfo {title} {Long-distance free-space quantum key distribution
  in daylight towards inter-satellite communication},\ }\href@noop {}
  {\bibfield  {journal} {\bibinfo  {journal} {Nature Photonics}\ }\textbf
  {\bibinfo {volume} {11}},\ \bibinfo {pages} {509} (\bibinfo {year}
  {2017})}\BibitemShut {NoStop}%
\bibitem [{\citenamefont {Strekalov}\ \emph {et~al.}(1995)\citenamefont
  {Strekalov}, \citenamefont {Sergienko}, \citenamefont {Klyshko},\ and\
  \citenamefont {Shih}}]{strekalov1995prl}%
  \BibitemOpen
  \bibfield  {author} {\bibinfo {author} {\bibfnamefont {D.~V.}\ \bibnamefont
  {Strekalov}}, \bibinfo {author} {\bibfnamefont {A.~V.}\ \bibnamefont
  {Sergienko}}, \bibinfo {author} {\bibfnamefont {D.~N.}\ \bibnamefont
  {Klyshko}},\ and\ \bibinfo {author} {\bibfnamefont {Y.~H.}\ \bibnamefont
  {Shih}},\ }\bibfield  {title} {\bibinfo {title} {Observation of two-photon
  ``ghost'' interference and diffraction},\ }\href
  {https://doi.org/10.1103/PhysRevLett.74.3600} {\bibfield  {journal} {\bibinfo
   {journal} {Phys. Rev. Lett.}\ }\textbf {\bibinfo {volume} {74}},\ \bibinfo
  {pages} {3600} (\bibinfo {year} {1995})}\BibitemShut {NoStop}%
\bibitem [{\citenamefont {Pittman}\ \emph {et~al.}(1995)\citenamefont
  {Pittman}, \citenamefont {Shih}, \citenamefont {Strekalov},\ and\
  \citenamefont {Sergienko}}]{pittman1995pra}%
  \BibitemOpen
  \bibfield  {author} {\bibinfo {author} {\bibfnamefont {T.}~\bibnamefont
  {Pittman}}, \bibinfo {author} {\bibfnamefont {Y.}~\bibnamefont {Shih}},
  \bibinfo {author} {\bibfnamefont {D.}~\bibnamefont {Strekalov}},\ and\
  \bibinfo {author} {\bibfnamefont {A.~V.}\ \bibnamefont {Sergienko}},\
  }\bibfield  {title} {\bibinfo {title} {Optical imaging by means of two-photon
  quantum entanglement},\ }\href@noop {} {\bibfield  {journal} {\bibinfo
  {journal} {Physical Review A}\ }\textbf {\bibinfo {volume} {52}},\ \bibinfo
  {pages} {R3429} (\bibinfo {year} {1995})}\BibitemShut {NoStop}%
\bibitem [{\citenamefont {Moreau}\ \emph {et~al.}(2018)\citenamefont {Moreau},
  \citenamefont {Toninelli}, \citenamefont {Gregory},\ and\ \citenamefont
  {Padgett}}]{moreau2018ghost}%
  \BibitemOpen
  \bibfield  {author} {\bibinfo {author} {\bibfnamefont {P.-A.}\ \bibnamefont
  {Moreau}}, \bibinfo {author} {\bibfnamefont {E.}~\bibnamefont {Toninelli}},
  \bibinfo {author} {\bibfnamefont {T.}~\bibnamefont {Gregory}},\ and\ \bibinfo
  {author} {\bibfnamefont {M.~J.}\ \bibnamefont {Padgett}},\ }\bibfield
  {title} {\bibinfo {title} {Ghost imaging using optical correlations},\
  }\href@noop {} {\bibfield  {journal} {\bibinfo  {journal} {Laser \& Photonics
  Reviews}\ }\textbf {\bibinfo {volume} {12}},\ \bibinfo {pages} {1700143}
  (\bibinfo {year} {2018})}\BibitemShut {NoStop}%
\bibitem [{\citenamefont {Brida}\ \emph {et~al.}(2010)\citenamefont {Brida},
  \citenamefont {Genovese},\ and\ \citenamefont
  {Ruo~Berchera}}]{brida2010natphot}%
  \BibitemOpen
  \bibfield  {author} {\bibinfo {author} {\bibfnamefont {G.}~\bibnamefont
  {Brida}}, \bibinfo {author} {\bibfnamefont {M.}~\bibnamefont {Genovese}},\
  and\ \bibinfo {author} {\bibfnamefont {I.}~\bibnamefont {Ruo~Berchera}},\
  }\bibfield  {title} {\bibinfo {title} {Experimental realization of
  sub-shot-noise quantum imaging},\ }\href@noop {} {\bibfield  {journal}
  {\bibinfo  {journal} {Nature Photonics}\ }\textbf {\bibinfo {volume} {4}},\
  \bibinfo {pages} {227} (\bibinfo {year} {2010})}\BibitemShut {NoStop}%
\bibitem [{\citenamefont {Blanchet}\ \emph {et~al.}(2008)\citenamefont
  {Blanchet}, \citenamefont {Devaux}, \citenamefont {Furfaro},\ and\
  \citenamefont {Lantz}}]{blanchet2008prl}%
  \BibitemOpen
  \bibfield  {author} {\bibinfo {author} {\bibfnamefont {J.-L.}\ \bibnamefont
  {Blanchet}}, \bibinfo {author} {\bibfnamefont {F.}~\bibnamefont {Devaux}},
  \bibinfo {author} {\bibfnamefont {L.}~\bibnamefont {Furfaro}},\ and\ \bibinfo
  {author} {\bibfnamefont {E.}~\bibnamefont {Lantz}},\ }\bibfield  {title}
  {\bibinfo {title} {Measurement of sub-shot-noise correlations of spatial
  fluctuations in the photon-counting regime},\ }\href@noop {} {\bibfield
  {journal} {\bibinfo  {journal} {Physical review letters}\ }\textbf {\bibinfo
  {volume} {101}},\ \bibinfo {pages} {233604} (\bibinfo {year}
  {2008})}\BibitemShut {NoStop}%
\bibitem [{\citenamefont {Giovannetti}\ \emph {et~al.}(2009)\citenamefont
  {Giovannetti}, \citenamefont {Lloyd}, \citenamefont {Maccone},\ and\
  \citenamefont {Shapiro}}]{giovannetti2009pra}%
  \BibitemOpen
  \bibfield  {author} {\bibinfo {author} {\bibfnamefont {V.}~\bibnamefont
  {Giovannetti}}, \bibinfo {author} {\bibfnamefont {S.}~\bibnamefont {Lloyd}},
  \bibinfo {author} {\bibfnamefont {L.}~\bibnamefont {Maccone}},\ and\ \bibinfo
  {author} {\bibfnamefont {J.~H.}\ \bibnamefont {Shapiro}},\ }\bibfield
  {title} {\bibinfo {title} {Sub-rayleigh-diffraction-bound quantum imaging},\
  }\href@noop {} {\bibfield  {journal} {\bibinfo  {journal} {Physical Review
  A}\ }\textbf {\bibinfo {volume} {79}},\ \bibinfo {pages} {013827} (\bibinfo
  {year} {2009})}\BibitemShut {NoStop}%
\bibitem [{\citenamefont {Untern{\"a}hrer}\ \emph {et~al.}(2018)\citenamefont
  {Untern{\"a}hrer}, \citenamefont {Bessire}, \citenamefont {Gasparini},
  \citenamefont {Perenzoni},\ and\ \citenamefont
  {Stefanov}}]{unternahrer2018optica}%
  \BibitemOpen
  \bibfield  {author} {\bibinfo {author} {\bibfnamefont {M.}~\bibnamefont
  {Untern{\"a}hrer}}, \bibinfo {author} {\bibfnamefont {B.}~\bibnamefont
  {Bessire}}, \bibinfo {author} {\bibfnamefont {L.}~\bibnamefont {Gasparini}},
  \bibinfo {author} {\bibfnamefont {M.}~\bibnamefont {Perenzoni}},\ and\
  \bibinfo {author} {\bibfnamefont {A.}~\bibnamefont {Stefanov}},\ }\bibfield
  {title} {\bibinfo {title} {Super-resolution quantum imaging at the heisenberg
  limit},\ }\href@noop {} {\bibfield  {journal} {\bibinfo  {journal} {Optica}\
  }\textbf {\bibinfo {volume} {5}},\ \bibinfo {pages} {1150} (\bibinfo {year}
  {2018})}\BibitemShut {NoStop}%
\bibitem [{\citenamefont {Santos}\ \emph {et~al.}(2022)\citenamefont {Santos},
  \citenamefont {Pertsch}, \citenamefont {Setzpfandt},\ and\ \citenamefont
  {Saravi}}]{santos2022prl}%
  \BibitemOpen
  \bibfield  {author} {\bibinfo {author} {\bibfnamefont {E.~A.}\ \bibnamefont
  {Santos}}, \bibinfo {author} {\bibfnamefont {T.}~\bibnamefont {Pertsch}},
  \bibinfo {author} {\bibfnamefont {F.}~\bibnamefont {Setzpfandt}},\ and\
  \bibinfo {author} {\bibfnamefont {S.}~\bibnamefont {Saravi}},\ }\bibfield
  {title} {\bibinfo {title} {Subdiffraction quantum imaging with undetected
  photons},\ }\href@noop {} {\bibfield  {journal} {\bibinfo  {journal}
  {Physical Review Letters}\ }\textbf {\bibinfo {volume} {128}},\ \bibinfo
  {pages} {173601} (\bibinfo {year} {2022})}\BibitemShut {NoStop}%
\bibitem [{\citenamefont {He}\ \emph {et~al.}(2023)\citenamefont {He},
  \citenamefont {Zhang}, \citenamefont {Tong}, \citenamefont {Li},\ and\
  \citenamefont {Wang}}]{he2023natcomm}%
  \BibitemOpen
  \bibfield  {author} {\bibinfo {author} {\bibfnamefont {Z.}~\bibnamefont
  {He}}, \bibinfo {author} {\bibfnamefont {Y.}~\bibnamefont {Zhang}}, \bibinfo
  {author} {\bibfnamefont {X.}~\bibnamefont {Tong}}, \bibinfo {author}
  {\bibfnamefont {L.}~\bibnamefont {Li}},\ and\ \bibinfo {author}
  {\bibfnamefont {L.~V.}\ \bibnamefont {Wang}},\ }\bibfield  {title} {\bibinfo
  {title} {Quantum microscopy of cells at the heisenberg limit},\ }\href@noop
  {} {\bibfield  {journal} {\bibinfo  {journal} {Nature Communications}\
  }\textbf {\bibinfo {volume} {14}},\ \bibinfo {pages} {2441} (\bibinfo {year}
  {2023})}\BibitemShut {NoStop}%
\bibitem [{\citenamefont {Lemos}\ \emph {et~al.}(2014)\citenamefont {Lemos},
  \citenamefont {Borish}, \citenamefont {Cole}, \citenamefont {Ramelow},
  \citenamefont {Lapkiewicz},\ and\ \citenamefont
  {Zeilinger}}]{lemos2014nature}%
  \BibitemOpen
  \bibfield  {author} {\bibinfo {author} {\bibfnamefont {G.~B.}\ \bibnamefont
  {Lemos}}, \bibinfo {author} {\bibfnamefont {V.}~\bibnamefont {Borish}},
  \bibinfo {author} {\bibfnamefont {G.~D.}\ \bibnamefont {Cole}}, \bibinfo
  {author} {\bibfnamefont {S.}~\bibnamefont {Ramelow}}, \bibinfo {author}
  {\bibfnamefont {R.}~\bibnamefont {Lapkiewicz}},\ and\ \bibinfo {author}
  {\bibfnamefont {A.}~\bibnamefont {Zeilinger}},\ }\bibfield  {title} {\bibinfo
  {title} {Quantum imaging with undetected photons},\ }\href@noop {} {\bibfield
   {journal} {\bibinfo  {journal} {Nature}\ }\textbf {\bibinfo {volume}
  {512}},\ \bibinfo {pages} {409} (\bibinfo {year} {2014})}\BibitemShut
  {NoStop}%
\bibitem [{\citenamefont {Fuenzalida}\ \emph {et~al.}(2023)\citenamefont
  {Fuenzalida}, \citenamefont {Gilaberte~Basset}, \citenamefont {T{\"o}pfer},
  \citenamefont {Torres},\ and\ \citenamefont
  {Gr{\"a}fe}}]{fuenzalida2023sciadv}%
  \BibitemOpen
  \bibfield  {author} {\bibinfo {author} {\bibfnamefont {J.}~\bibnamefont
  {Fuenzalida}}, \bibinfo {author} {\bibfnamefont {M.}~\bibnamefont
  {Gilaberte~Basset}}, \bibinfo {author} {\bibfnamefont {S.}~\bibnamefont
  {T{\"o}pfer}}, \bibinfo {author} {\bibfnamefont {J.~P.}\ \bibnamefont
  {Torres}},\ and\ \bibinfo {author} {\bibfnamefont {M.}~\bibnamefont
  {Gr{\"a}fe}},\ }\bibfield  {title} {\bibinfo {title} {Experimental quantum
  imaging distillation with undetected light},\ }\href@noop {} {\bibfield
  {journal} {\bibinfo  {journal} {Science Advances}\ }\textbf {\bibinfo
  {volume} {9}},\ \bibinfo {pages} {eadg9573} (\bibinfo {year}
  {2023})}\BibitemShut {NoStop}%
\bibitem [{\citenamefont {Defienne}\ \emph {et~al.}(2019)\citenamefont
  {Defienne}, \citenamefont {Reichert}, \citenamefont {Fleischer},\ and\
  \citenamefont {Faccio}}]{defienne2019sciadv}%
  \BibitemOpen
  \bibfield  {author} {\bibinfo {author} {\bibfnamefont {H.}~\bibnamefont
  {Defienne}}, \bibinfo {author} {\bibfnamefont {M.}~\bibnamefont {Reichert}},
  \bibinfo {author} {\bibfnamefont {J.~W.}\ \bibnamefont {Fleischer}},\ and\
  \bibinfo {author} {\bibfnamefont {D.}~\bibnamefont {Faccio}},\ }\bibfield
  {title} {\bibinfo {title} {Quantum image distillation},\ }\href@noop {}
  {\bibfield  {journal} {\bibinfo  {journal} {Science advances}\ }\textbf
  {\bibinfo {volume} {5}},\ \bibinfo {pages} {eaax0307} (\bibinfo {year}
  {2019})}\BibitemShut {NoStop}%
\bibitem [{\citenamefont {Cameron}\ \emph {et~al.}(2024)\citenamefont
  {Cameron}, \citenamefont {Courme}, \citenamefont {Verni{\`e}re},
  \citenamefont {Pandya}, \citenamefont {Faccio},\ and\ \citenamefont
  {Defienne}}]{cameron2024science}%
  \BibitemOpen
  \bibfield  {author} {\bibinfo {author} {\bibfnamefont {P.}~\bibnamefont
  {Cameron}}, \bibinfo {author} {\bibfnamefont {B.}~\bibnamefont {Courme}},
  \bibinfo {author} {\bibfnamefont {C.}~\bibnamefont {Verni{\`e}re}}, \bibinfo
  {author} {\bibfnamefont {R.}~\bibnamefont {Pandya}}, \bibinfo {author}
  {\bibfnamefont {D.}~\bibnamefont {Faccio}},\ and\ \bibinfo {author}
  {\bibfnamefont {H.}~\bibnamefont {Defienne}},\ }\bibfield  {title} {\bibinfo
  {title} {Adaptive optical imaging with entangled photons},\ }\href@noop {}
  {\bibfield  {journal} {\bibinfo  {journal} {Science}\ }\textbf {\bibinfo
  {volume} {383}},\ \bibinfo {pages} {1142} (\bibinfo {year}
  {2024})}\BibitemShut {NoStop}%
\bibitem [{\citenamefont {Defienne}\ \emph {et~al.}(2021)\citenamefont
  {Defienne}, \citenamefont {Ndagano}, \citenamefont {Lyons},\ and\
  \citenamefont {Faccio}}]{defienne2021natphy}%
  \BibitemOpen
  \bibfield  {author} {\bibinfo {author} {\bibfnamefont {H.}~\bibnamefont
  {Defienne}}, \bibinfo {author} {\bibfnamefont {B.}~\bibnamefont {Ndagano}},
  \bibinfo {author} {\bibfnamefont {A.}~\bibnamefont {Lyons}},\ and\ \bibinfo
  {author} {\bibfnamefont {D.}~\bibnamefont {Faccio}},\ }\bibfield  {title}
  {\bibinfo {title} {Polarization entanglement-enabled quantum holography},\
  }\href@noop {} {\bibfield  {journal} {\bibinfo  {journal} {Nature Physics}\
  }\textbf {\bibinfo {volume} {17}},\ \bibinfo {pages} {591} (\bibinfo {year}
  {2021})}\BibitemShut {NoStop}%
\bibitem [{\citenamefont {Nirala}\ \emph {et~al.}(2023)\citenamefont {Nirala},
  \citenamefont {Pradyumna}, \citenamefont {Kumar},\ and\ \citenamefont
  {Marino}}]{nirala2023sciadv}%
  \BibitemOpen
  \bibfield  {author} {\bibinfo {author} {\bibfnamefont {G.}~\bibnamefont
  {Nirala}}, \bibinfo {author} {\bibfnamefont {S.~T.}\ \bibnamefont
  {Pradyumna}}, \bibinfo {author} {\bibfnamefont {A.}~\bibnamefont {Kumar}},\
  and\ \bibinfo {author} {\bibfnamefont {A.~M.}\ \bibnamefont {Marino}},\
  }\bibfield  {title} {\bibinfo {title} {Information encoding in the spatial
  correlations of entangled twin beams},\ }\href@noop {} {\bibfield  {journal}
  {\bibinfo  {journal} {Science Advances}\ }\textbf {\bibinfo {volume} {9}},\
  \bibinfo {pages} {eadf9161} (\bibinfo {year} {2023})}\BibitemShut {NoStop}%
\bibitem [{\citenamefont {Genovese}(2016)}]{genovese2016jopt}%
  \BibitemOpen
  \bibfield  {author} {\bibinfo {author} {\bibfnamefont {M.}~\bibnamefont
  {Genovese}},\ }\bibfield  {title} {\bibinfo {title} {Real applications of
  quantum imaging},\ }\href@noop {} {\bibfield  {journal} {\bibinfo  {journal}
  {Journal of Optics}\ }\textbf {\bibinfo {volume} {18}},\ \bibinfo {pages}
  {073002} (\bibinfo {year} {2016})}\BibitemShut {NoStop}%
\bibitem [{\citenamefont {Moreau}\ \emph {et~al.}(2019)\citenamefont {Moreau},
  \citenamefont {Toninelli}, \citenamefont {Gregory},\ and\ \citenamefont
  {Padgett}}]{moreau2019natrevphy}%
  \BibitemOpen
  \bibfield  {author} {\bibinfo {author} {\bibfnamefont {P.-A.}\ \bibnamefont
  {Moreau}}, \bibinfo {author} {\bibfnamefont {E.}~\bibnamefont {Toninelli}},
  \bibinfo {author} {\bibfnamefont {T.}~\bibnamefont {Gregory}},\ and\ \bibinfo
  {author} {\bibfnamefont {M.~J.}\ \bibnamefont {Padgett}},\ }\bibfield
  {title} {\bibinfo {title} {Imaging with quantum states of light},\
  }\href@noop {} {\bibfield  {journal} {\bibinfo  {journal} {Nature Reviews
  Physics}\ }\textbf {\bibinfo {volume} {1}},\ \bibinfo {pages} {367} (\bibinfo
  {year} {2019})}\BibitemShut {NoStop}%
\bibitem [{\citenamefont {Erkmen}\ and\ \citenamefont
  {Shapiro}(2010)}]{erkmen2010advoptphot}%
  \BibitemOpen
  \bibfield  {author} {\bibinfo {author} {\bibfnamefont {B.~I.}\ \bibnamefont
  {Erkmen}}\ and\ \bibinfo {author} {\bibfnamefont {J.~H.}\ \bibnamefont
  {Shapiro}},\ }\bibfield  {title} {\bibinfo {title} {Ghost imaging: from
  quantum to classical to computational},\ }\href@noop {} {\bibfield  {journal}
  {\bibinfo  {journal} {Advances in Optics and Photonics}\ }\textbf {\bibinfo
  {volume} {2}},\ \bibinfo {pages} {405} (\bibinfo {year} {2010})}\BibitemShut
  {NoStop}%
\bibitem [{\citenamefont {Padgett}\ and\ \citenamefont
  {Boyd}(2017)}]{padgett2017philtransrsa}%
  \BibitemOpen
  \bibfield  {author} {\bibinfo {author} {\bibfnamefont {M.~J.}\ \bibnamefont
  {Padgett}}\ and\ \bibinfo {author} {\bibfnamefont {R.~W.}\ \bibnamefont
  {Boyd}},\ }\bibfield  {title} {\bibinfo {title} {An introduction to ghost
  imaging: quantum and classical},\ }\href@noop {} {\bibfield  {journal}
  {\bibinfo  {journal} {Philosophical Transactions of the Royal Society A:
  Mathematical, Physical and Engineering Sciences}\ }\textbf {\bibinfo {volume}
  {375}},\ \bibinfo {pages} {20160233} (\bibinfo {year} {2017})}\BibitemShut
  {NoStop}%
\bibitem [{\citenamefont {Karan}\ \emph {et~al.}(2020)\citenamefont {Karan},
  \citenamefont {Aarav}, \citenamefont {Bharadhwaj}, \citenamefont {Taneja},
  \citenamefont {De}, \citenamefont {Kulkarni}, \citenamefont {Meher},\ and\
  \citenamefont {Jha}}]{karan2020jopt}%
  \BibitemOpen
  \bibfield  {author} {\bibinfo {author} {\bibfnamefont {S.}~\bibnamefont
  {Karan}}, \bibinfo {author} {\bibfnamefont {S.}~\bibnamefont {Aarav}},
  \bibinfo {author} {\bibfnamefont {H.}~\bibnamefont {Bharadhwaj}}, \bibinfo
  {author} {\bibfnamefont {L.}~\bibnamefont {Taneja}}, \bibinfo {author}
  {\bibfnamefont {A.}~\bibnamefont {De}}, \bibinfo {author} {\bibfnamefont
  {G.}~\bibnamefont {Kulkarni}}, \bibinfo {author} {\bibfnamefont
  {N.}~\bibnamefont {Meher}},\ and\ \bibinfo {author} {\bibfnamefont {A.~K.}\
  \bibnamefont {Jha}},\ }\bibfield  {title} {\bibinfo {title} {Phase matching
  in $\beta$-barium borate crystals for spontaneous parametric
  down-conversion},\ }\href@noop {} {\bibfield  {journal} {\bibinfo  {journal}
  {Journal of Optics}\ }\textbf {\bibinfo {volume} {22}},\ \bibinfo {pages}
  {083501} (\bibinfo {year} {2020})}\BibitemShut {NoStop}%
\bibitem [{\citenamefont {Hong}\ and\ \citenamefont
  {Mandel}(1985)}]{hong1985pra}%
  \BibitemOpen
  \bibfield  {author} {\bibinfo {author} {\bibfnamefont {C.}~\bibnamefont
  {Hong}}\ and\ \bibinfo {author} {\bibfnamefont {L.}~\bibnamefont {Mandel}},\
  }\bibfield  {title} {\bibinfo {title} {Theory of parametric frequency down
  conversion of light},\ }\href@noop {} {\bibfield  {journal} {\bibinfo
  {journal} {Physical Review A}\ }\textbf {\bibinfo {volume} {31}},\ \bibinfo
  {pages} {2409} (\bibinfo {year} {1985})}\BibitemShut {NoStop}%
\bibitem [{\citenamefont {Kwiat}\ \emph {et~al.}(1995)\citenamefont {Kwiat},
  \citenamefont {Mattle}, \citenamefont {Weinfurter}, \citenamefont
  {Zeilinger}, \citenamefont {Sergienko},\ and\ \citenamefont
  {Shih}}]{kwiat1995prl}%
  \BibitemOpen
  \bibfield  {author} {\bibinfo {author} {\bibfnamefont {P.~G.}\ \bibnamefont
  {Kwiat}}, \bibinfo {author} {\bibfnamefont {K.}~\bibnamefont {Mattle}},
  \bibinfo {author} {\bibfnamefont {H.}~\bibnamefont {Weinfurter}}, \bibinfo
  {author} {\bibfnamefont {A.}~\bibnamefont {Zeilinger}}, \bibinfo {author}
  {\bibfnamefont {A.~V.}\ \bibnamefont {Sergienko}},\ and\ \bibinfo {author}
  {\bibfnamefont {Y.}~\bibnamefont {Shih}},\ }\bibfield  {title} {\bibinfo
  {title} {New high-intensity source of polarization-entangled photon pairs},\
  }\href@noop {} {\bibfield  {journal} {\bibinfo  {journal} {Physical Review
  Letters}\ }\textbf {\bibinfo {volume} {75}},\ \bibinfo {pages} {4337}
  (\bibinfo {year} {1995})}\BibitemShut {NoStop}%
\bibitem [{\citenamefont {Leach}\ \emph {et~al.}(2010)\citenamefont {Leach},
  \citenamefont {Jack}, \citenamefont {Romero}, \citenamefont {Jha},
  \citenamefont {Yao}, \citenamefont {Franke-Arnold}, \citenamefont {Ireland},
  \citenamefont {Boyd}, \citenamefont {Barnett},\ and\ \citenamefont
  {Padgett}}]{leach2010science}%
  \BibitemOpen
  \bibfield  {author} {\bibinfo {author} {\bibfnamefont {J.}~\bibnamefont
  {Leach}}, \bibinfo {author} {\bibfnamefont {B.}~\bibnamefont {Jack}},
  \bibinfo {author} {\bibfnamefont {J.}~\bibnamefont {Romero}}, \bibinfo
  {author} {\bibfnamefont {A.~K.}\ \bibnamefont {Jha}}, \bibinfo {author}
  {\bibfnamefont {A.~M.}\ \bibnamefont {Yao}}, \bibinfo {author} {\bibfnamefont
  {S.}~\bibnamefont {Franke-Arnold}}, \bibinfo {author} {\bibfnamefont {D.~G.}\
  \bibnamefont {Ireland}}, \bibinfo {author} {\bibfnamefont {R.~W.}\
  \bibnamefont {Boyd}}, \bibinfo {author} {\bibfnamefont {S.~M.}\ \bibnamefont
  {Barnett}},\ and\ \bibinfo {author} {\bibfnamefont {M.~J.}\ \bibnamefont
  {Padgett}},\ }\bibfield  {title} {\bibinfo {title} {Quantum correlations in
  optical angle--orbital angular momentum variables},\ }\href@noop {}
  {\bibfield  {journal} {\bibinfo  {journal} {Science}\ }\textbf {\bibinfo
  {volume} {329}},\ \bibinfo {pages} {662} (\bibinfo {year}
  {2010})}\BibitemShut {NoStop}%
\bibitem [{\citenamefont {Thew}\ \emph {et~al.}(2004)\citenamefont {Thew},
  \citenamefont {Acin}, \citenamefont {Zbinden},\ and\ \citenamefont
  {Gisin}}]{thew2004prl}%
  \BibitemOpen
  \bibfield  {author} {\bibinfo {author} {\bibfnamefont {R.~T.}\ \bibnamefont
  {Thew}}, \bibinfo {author} {\bibfnamefont {A.}~\bibnamefont {Acin}}, \bibinfo
  {author} {\bibfnamefont {H.}~\bibnamefont {Zbinden}},\ and\ \bibinfo {author}
  {\bibfnamefont {N.}~\bibnamefont {Gisin}},\ }\bibfield  {title} {\bibinfo
  {title} {Bell-type test of energy-time entangled qutrits},\ }\href@noop {}
  {\bibfield  {journal} {\bibinfo  {journal} {Physical review letters}\
  }\textbf {\bibinfo {volume} {93}},\ \bibinfo {pages} {010503} (\bibinfo
  {year} {2004})}\BibitemShut {NoStop}%
\bibitem [{\citenamefont {O’Sullivan-Hale}\ \emph {et~al.}(2005)\citenamefont
  {O’Sullivan-Hale}, \citenamefont {Khan}, \citenamefont {Boyd},\ and\
  \citenamefont {Howell}}]{o2005prl}%
  \BibitemOpen
  \bibfield  {author} {\bibinfo {author} {\bibfnamefont {M.~N.}\ \bibnamefont
  {O’Sullivan-Hale}}, \bibinfo {author} {\bibfnamefont {I.~A.}\ \bibnamefont
  {Khan}}, \bibinfo {author} {\bibfnamefont {R.~W.}\ \bibnamefont {Boyd}},\
  and\ \bibinfo {author} {\bibfnamefont {J.~C.}\ \bibnamefont {Howell}},\
  }\bibfield  {title} {\bibinfo {title} {Pixel entanglement: experimental
  realization of optically entangled d= 3 and d= 6 qudits},\ }\href@noop {}
  {\bibfield  {journal} {\bibinfo  {journal} {Physical review letters}\
  }\textbf {\bibinfo {volume} {94}},\ \bibinfo {pages} {220501} (\bibinfo
  {year} {2005})}\BibitemShut {NoStop}%
\bibitem [{\citenamefont {Walborn}\ \emph {et~al.}(2010)\citenamefont
  {Walborn}, \citenamefont {Monken}, \citenamefont {P{\'a}dua},\ and\
  \citenamefont {Ribeiro}}]{walborn2010phyrep}%
  \BibitemOpen
  \bibfield  {author} {\bibinfo {author} {\bibfnamefont {S.~P.}\ \bibnamefont
  {Walborn}}, \bibinfo {author} {\bibfnamefont {C.}~\bibnamefont {Monken}},
  \bibinfo {author} {\bibfnamefont {S.}~\bibnamefont {P{\'a}dua}},\ and\
  \bibinfo {author} {\bibfnamefont {P.~S.}\ \bibnamefont {Ribeiro}},\
  }\bibfield  {title} {\bibinfo {title} {Spatial correlations in parametric
  down-conversion},\ }\href@noop {} {\bibfield  {journal} {\bibinfo  {journal}
  {Physics Reports}\ }\textbf {\bibinfo {volume} {495}},\ \bibinfo {pages} {87}
  (\bibinfo {year} {2010})}\BibitemShut {NoStop}%
\bibitem [{\citenamefont {Bhattacharjee}\ \emph
  {et~al.}(2022{\natexlab{a}})\citenamefont {Bhattacharjee}, \citenamefont
  {Meher},\ and\ \citenamefont {Jha}}]{bhattacharjee2022njp}%
  \BibitemOpen
  \bibfield  {author} {\bibinfo {author} {\bibfnamefont {A.}~\bibnamefont
  {Bhattacharjee}}, \bibinfo {author} {\bibfnamefont {N.}~\bibnamefont
  {Meher}},\ and\ \bibinfo {author} {\bibfnamefont {A.~K.}\ \bibnamefont
  {Jha}},\ }\bibfield  {title} {\bibinfo {title} {Measurement of two-photon
  position--momentum einstein--podolsky--rosen correlations through
  single-photon intensity measurements},\ }\href@noop {} {\bibfield  {journal}
  {\bibinfo  {journal} {New Journal of Physics}\ }\textbf {\bibinfo {volume}
  {24}},\ \bibinfo {pages} {053033} (\bibinfo {year}
  {2022}{\natexlab{a}})}\BibitemShut {NoStop}%
\bibitem [{\citenamefont {Berta}\ \emph {et~al.}(2010)\citenamefont {Berta},
  \citenamefont {Christandl}, \citenamefont {Colbeck}, \citenamefont {Renes},\
  and\ \citenamefont {Renner}}]{berta2010natphy}%
  \BibitemOpen
  \bibfield  {author} {\bibinfo {author} {\bibfnamefont {M.}~\bibnamefont
  {Berta}}, \bibinfo {author} {\bibfnamefont {M.}~\bibnamefont {Christandl}},
  \bibinfo {author} {\bibfnamefont {R.}~\bibnamefont {Colbeck}}, \bibinfo
  {author} {\bibfnamefont {J.~M.}\ \bibnamefont {Renes}},\ and\ \bibinfo
  {author} {\bibfnamefont {R.}~\bibnamefont {Renner}},\ }\bibfield  {title}
  {\bibinfo {title} {The uncertainty principle in the presence of quantum
  memory},\ }\href@noop {} {\bibfield  {journal} {\bibinfo  {journal} {Nature
  Physics}\ }\textbf {\bibinfo {volume} {6}},\ \bibinfo {pages} {659} (\bibinfo
  {year} {2010})}\BibitemShut {NoStop}%
\bibitem [{\citenamefont {Howell}\ \emph {et~al.}(2004)\citenamefont {Howell},
  \citenamefont {Bennink}, \citenamefont {Bentley},\ and\ \citenamefont
  {Boyd}}]{howell2004prl}%
  \BibitemOpen
  \bibfield  {author} {\bibinfo {author} {\bibfnamefont {J.~C.}\ \bibnamefont
  {Howell}}, \bibinfo {author} {\bibfnamefont {R.~S.}\ \bibnamefont {Bennink}},
  \bibinfo {author} {\bibfnamefont {S.~J.}\ \bibnamefont {Bentley}},\ and\
  \bibinfo {author} {\bibfnamefont {R.~W.}\ \bibnamefont {Boyd}},\ }\bibfield
  {title} {\bibinfo {title} {Realization of the einstein-podolsky-rosen paradox
  using momentum-and position-entangled photons from spontaneous parametric
  down conversion},\ }\href@noop {} {\bibfield  {journal} {\bibinfo  {journal}
  {Physical Review Letters}\ }\textbf {\bibinfo {volume} {92}},\ \bibinfo
  {pages} {210403} (\bibinfo {year} {2004})}\BibitemShut {NoStop}%
\bibitem [{\citenamefont {Bhattacharjee}\ \emph
  {et~al.}(2022{\natexlab{b}})\citenamefont {Bhattacharjee}, \citenamefont
  {Joshi}, \citenamefont {Karan}, \citenamefont {Leach},\ and\ \citenamefont
  {Jha}}]{bhattacharjee2022sciadv}%
  \BibitemOpen
  \bibfield  {author} {\bibinfo {author} {\bibfnamefont {A.}~\bibnamefont
  {Bhattacharjee}}, \bibinfo {author} {\bibfnamefont {M.~K.}\ \bibnamefont
  {Joshi}}, \bibinfo {author} {\bibfnamefont {S.}~\bibnamefont {Karan}},
  \bibinfo {author} {\bibfnamefont {J.}~\bibnamefont {Leach}},\ and\ \bibinfo
  {author} {\bibfnamefont {A.~K.}\ \bibnamefont {Jha}},\ }\bibfield  {title}
  {\bibinfo {title} {Propagation-induced revival of entanglement in the
  angle-oam bases},\ }\href@noop {} {\bibfield  {journal} {\bibinfo  {journal}
  {Science Advances}\ }\textbf {\bibinfo {volume} {8}},\ \bibinfo {pages}
  {eabn7876} (\bibinfo {year} {2022}{\natexlab{b}})}\BibitemShut {NoStop}%
\bibitem [{\citenamefont {Boucher}\ \emph {et~al.}(2021)\citenamefont
  {Boucher}, \citenamefont {Defienne},\ and\ \citenamefont
  {Gigan}}]{boucher2021optlett}%
  \BibitemOpen
  \bibfield  {author} {\bibinfo {author} {\bibfnamefont {P.}~\bibnamefont
  {Boucher}}, \bibinfo {author} {\bibfnamefont {H.}~\bibnamefont {Defienne}},\
  and\ \bibinfo {author} {\bibfnamefont {S.}~\bibnamefont {Gigan}},\ }\bibfield
   {title} {\bibinfo {title} {Engineering spatial correlations of entangled
  photon pairs by pump beam shaping},\ }\href@noop {} {\bibfield  {journal}
  {\bibinfo  {journal} {Optics Letters}\ }\textbf {\bibinfo {volume} {46}},\
  \bibinfo {pages} {4200} (\bibinfo {year} {2021})}\BibitemShut {NoStop}%
\bibitem [{\citenamefont {Valencia}\ \emph {et~al.}(2007)\citenamefont
  {Valencia}, \citenamefont {Cer{\'e}}, \citenamefont {Shi}, \citenamefont
  {Molina-Terriza},\ and\ \citenamefont {Torres}}]{valencia2007prl}%
  \BibitemOpen
  \bibfield  {author} {\bibinfo {author} {\bibfnamefont {A.}~\bibnamefont
  {Valencia}}, \bibinfo {author} {\bibfnamefont {A.}~\bibnamefont {Cer{\'e}}},
  \bibinfo {author} {\bibfnamefont {X.}~\bibnamefont {Shi}}, \bibinfo {author}
  {\bibfnamefont {G.}~\bibnamefont {Molina-Terriza}},\ and\ \bibinfo {author}
  {\bibfnamefont {J.~P.}\ \bibnamefont {Torres}},\ }\bibfield  {title}
  {\bibinfo {title} {Shaping the waveform of entangled photons},\ }\href@noop
  {} {\bibfield  {journal} {\bibinfo  {journal} {Physical review letters}\
  }\textbf {\bibinfo {volume} {99}},\ \bibinfo {pages} {243601} (\bibinfo
  {year} {2007})}\BibitemShut {NoStop}%
\bibitem [{\citenamefont {Monken}\ \emph {et~al.}(1998)\citenamefont {Monken},
  \citenamefont {Ribeiro},\ and\ \citenamefont {P{\'a}dua}}]{monken1998pra}%
  \BibitemOpen
  \bibfield  {author} {\bibinfo {author} {\bibfnamefont {C.~H.}\ \bibnamefont
  {Monken}}, \bibinfo {author} {\bibfnamefont {P.~S.}\ \bibnamefont
  {Ribeiro}},\ and\ \bibinfo {author} {\bibfnamefont {S.}~\bibnamefont
  {P{\'a}dua}},\ }\bibfield  {title} {\bibinfo {title} {Transfer of angular
  spectrum and image formation in spontaneous parametric down-conversion},\
  }\href@noop {} {\bibfield  {journal} {\bibinfo  {journal} {Physical Review
  A}\ }\textbf {\bibinfo {volume} {57}},\ \bibinfo {pages} {3123} (\bibinfo
  {year} {1998})}\BibitemShut {NoStop}%
\bibitem [{\citenamefont {Zhang}\ \emph {et~al.}(2019)\citenamefont {Zhang},
  \citenamefont {Fickler}, \citenamefont {Giese}, \citenamefont {Chen},\ and\
  \citenamefont {Boyd}}]{zhang2019optexp}%
  \BibitemOpen
  \bibfield  {author} {\bibinfo {author} {\bibfnamefont {W.}~\bibnamefont
  {Zhang}}, \bibinfo {author} {\bibfnamefont {R.}~\bibnamefont {Fickler}},
  \bibinfo {author} {\bibfnamefont {E.}~\bibnamefont {Giese}}, \bibinfo
  {author} {\bibfnamefont {L.}~\bibnamefont {Chen}},\ and\ \bibinfo {author}
  {\bibfnamefont {R.~W.}\ \bibnamefont {Boyd}},\ }\bibfield  {title} {\bibinfo
  {title} {Influence of pump coherence on the generation of position-momentum
  entanglement in optical parametric down-conversion},\ }\href@noop {}
  {\bibfield  {journal} {\bibinfo  {journal} {Optics express}\ }\textbf
  {\bibinfo {volume} {27}},\ \bibinfo {pages} {20745} (\bibinfo {year}
  {2019})}\BibitemShut {NoStop}%
\bibitem [{\citenamefont {Zia}\ \emph {et~al.}(2023)\citenamefont {Zia},
  \citenamefont {Dehghan}, \citenamefont {D’Errico}, \citenamefont
  {Sciarrino},\ and\ \citenamefont {Karimi}}]{zia2023natphot}%
  \BibitemOpen
  \bibfield  {author} {\bibinfo {author} {\bibfnamefont {D.}~\bibnamefont
  {Zia}}, \bibinfo {author} {\bibfnamefont {N.}~\bibnamefont {Dehghan}},
  \bibinfo {author} {\bibfnamefont {A.}~\bibnamefont {D’Errico}}, \bibinfo
  {author} {\bibfnamefont {F.}~\bibnamefont {Sciarrino}},\ and\ \bibinfo
  {author} {\bibfnamefont {E.}~\bibnamefont {Karimi}},\ }\bibfield  {title}
  {\bibinfo {title} {Interferometric imaging of amplitude and phase of spatial
  biphoton states},\ }\href@noop {} {\bibfield  {journal} {\bibinfo  {journal}
  {Nature Photonics}\ }\textbf {\bibinfo {volume} {17}},\ \bibinfo {pages}
  {1009} (\bibinfo {year} {2023})}\BibitemShut {NoStop}%
\bibitem [{\citenamefont {Prevedel}\ \emph {et~al.}(2011)\citenamefont
  {Prevedel}, \citenamefont {Hamel}, \citenamefont {Colbeck}, \citenamefont
  {Fisher},\ and\ \citenamefont {Resch}}]{prevedel2011natphy}%
  \BibitemOpen
  \bibfield  {author} {\bibinfo {author} {\bibfnamefont {R.}~\bibnamefont
  {Prevedel}}, \bibinfo {author} {\bibfnamefont {D.~R.}\ \bibnamefont {Hamel}},
  \bibinfo {author} {\bibfnamefont {R.}~\bibnamefont {Colbeck}}, \bibinfo
  {author} {\bibfnamefont {K.}~\bibnamefont {Fisher}},\ and\ \bibinfo {author}
  {\bibfnamefont {K.~J.}\ \bibnamefont {Resch}},\ }\bibfield  {title} {\bibinfo
  {title} {Experimental investigation of the uncertainty principle in the
  presence of quantum memory and its application to witnessing entanglement},\
  }\href@noop {} {\bibfield  {journal} {\bibinfo  {journal} {Nature Physics}\
  }\textbf {\bibinfo {volume} {7}},\ \bibinfo {pages} {757} (\bibinfo {year}
  {2011})}\BibitemShut {NoStop}%
\bibitem [{\citenamefont {Schneeloch}\ \emph
  {et~al.}(2019{\natexlab{a}})\citenamefont {Schneeloch}, \citenamefont
  {Tison}, \citenamefont {Fanto}, \citenamefont {Alsing},\ and\ \citenamefont
  {Howland}}]{schneeloch2019natcomm}%
  \BibitemOpen
  \bibfield  {author} {\bibinfo {author} {\bibfnamefont {J.}~\bibnamefont
  {Schneeloch}}, \bibinfo {author} {\bibfnamefont {C.~C.}\ \bibnamefont
  {Tison}}, \bibinfo {author} {\bibfnamefont {M.~L.}\ \bibnamefont {Fanto}},
  \bibinfo {author} {\bibfnamefont {P.~M.}\ \bibnamefont {Alsing}},\ and\
  \bibinfo {author} {\bibfnamefont {G.~A.}\ \bibnamefont {Howland}},\
  }\bibfield  {title} {\bibinfo {title} {Quantifying entanglement in a
  68-billion-dimensional quantum state space},\ }\href@noop {} {\bibfield
  {journal} {\bibinfo  {journal} {Nature communications}\ }\textbf {\bibinfo
  {volume} {10}},\ \bibinfo {pages} {1} (\bibinfo {year}
  {2019}{\natexlab{a}})}\BibitemShut {NoStop}%
\bibitem [{\citenamefont {Schneeloch}\ and\ \citenamefont
  {Howland}(2018)}]{schneeloch2018pra}%
  \BibitemOpen
  \bibfield  {author} {\bibinfo {author} {\bibfnamefont {J.}~\bibnamefont
  {Schneeloch}}\ and\ \bibinfo {author} {\bibfnamefont {G.~A.}\ \bibnamefont
  {Howland}},\ }\bibfield  {title} {\bibinfo {title} {Quantifying
  high-dimensional entanglement with einstein-podolsky-rosen correlations},\
  }\href@noop {} {\bibfield  {journal} {\bibinfo  {journal} {Physical Review
  A}\ }\textbf {\bibinfo {volume} {97}},\ \bibinfo {pages} {042338} (\bibinfo
  {year} {2018})}\BibitemShut {NoStop}%
\bibitem [{\citenamefont {Edgar}\ \emph {et~al.}(2012)\citenamefont {Edgar},
  \citenamefont {Tasca}, \citenamefont {Izdebski}, \citenamefont {Warburton},
  \citenamefont {Leach}, \citenamefont {Agnew}, \citenamefont {Buller},
  \citenamefont {Boyd},\ and\ \citenamefont {Padgett}}]{edgar2012natcomm}%
  \BibitemOpen
  \bibfield  {author} {\bibinfo {author} {\bibfnamefont {M.~P.}\ \bibnamefont
  {Edgar}}, \bibinfo {author} {\bibfnamefont {D.~S.}\ \bibnamefont {Tasca}},
  \bibinfo {author} {\bibfnamefont {F.}~\bibnamefont {Izdebski}}, \bibinfo
  {author} {\bibfnamefont {R.~E.}\ \bibnamefont {Warburton}}, \bibinfo {author}
  {\bibfnamefont {J.}~\bibnamefont {Leach}}, \bibinfo {author} {\bibfnamefont
  {M.}~\bibnamefont {Agnew}}, \bibinfo {author} {\bibfnamefont {G.~S.}\
  \bibnamefont {Buller}}, \bibinfo {author} {\bibfnamefont {R.~W.}\
  \bibnamefont {Boyd}},\ and\ \bibinfo {author} {\bibfnamefont {M.~J.}\
  \bibnamefont {Padgett}},\ }\bibfield  {title} {\bibinfo {title} {Imaging
  high-dimensional spatial entanglement with a camera},\ }\href@noop {}
  {\bibfield  {journal} {\bibinfo  {journal} {Nature communications}\ }\textbf
  {\bibinfo {volume} {3}},\ \bibinfo {pages} {1} (\bibinfo {year}
  {2012})}\BibitemShut {NoStop}%
\bibitem [{\citenamefont {Defienne}\ \emph {et~al.}(2018)\citenamefont
  {Defienne}, \citenamefont {Reichert},\ and\ \citenamefont
  {Fleischer}}]{defienne2018prl}%
  \BibitemOpen
  \bibfield  {author} {\bibinfo {author} {\bibfnamefont {H.}~\bibnamefont
  {Defienne}}, \bibinfo {author} {\bibfnamefont {M.}~\bibnamefont {Reichert}},\
  and\ \bibinfo {author} {\bibfnamefont {J.~W.}\ \bibnamefont {Fleischer}},\
  }\bibfield  {title} {\bibinfo {title} {General model of photon-pair detection
  with an image sensor},\ }\href@noop {} {\bibfield  {journal} {\bibinfo
  {journal} {Physical review letters}\ }\textbf {\bibinfo {volume} {120}},\
  \bibinfo {pages} {203604} (\bibinfo {year} {2018})}\BibitemShut {NoStop}%
\bibitem [{\citenamefont {Reichert}\ \emph {et~al.}(2018)\citenamefont
  {Reichert}, \citenamefont {Defienne},\ and\ \citenamefont
  {Fleischer}}]{reichert2018scirep}%
  \BibitemOpen
  \bibfield  {author} {\bibinfo {author} {\bibfnamefont {M.}~\bibnamefont
  {Reichert}}, \bibinfo {author} {\bibfnamefont {H.}~\bibnamefont {Defienne}},\
  and\ \bibinfo {author} {\bibfnamefont {J.~W.}\ \bibnamefont {Fleischer}},\
  }\bibfield  {title} {\bibinfo {title} {Massively parallel coincidence
  counting of high-dimensional entangled states},\ }\href@noop {} {\bibfield
  {journal} {\bibinfo  {journal} {Scientific reports}\ }\textbf {\bibinfo
  {volume} {8}},\ \bibinfo {pages} {7925} (\bibinfo {year} {2018})}\BibitemShut
  {NoStop}%
\bibitem [{\citenamefont {Ndagano}\ \emph {et~al.}(2020)\citenamefont
  {Ndagano}, \citenamefont {Defienne}, \citenamefont {Lyons}, \citenamefont
  {Starshynov}, \citenamefont {Villa}, \citenamefont {Tisa},\ and\
  \citenamefont {Faccio}}]{ndagano2020npjqinfo}%
  \BibitemOpen
  \bibfield  {author} {\bibinfo {author} {\bibfnamefont {B.}~\bibnamefont
  {Ndagano}}, \bibinfo {author} {\bibfnamefont {H.}~\bibnamefont {Defienne}},
  \bibinfo {author} {\bibfnamefont {A.}~\bibnamefont {Lyons}}, \bibinfo
  {author} {\bibfnamefont {I.}~\bibnamefont {Starshynov}}, \bibinfo {author}
  {\bibfnamefont {F.}~\bibnamefont {Villa}}, \bibinfo {author} {\bibfnamefont
  {S.}~\bibnamefont {Tisa}},\ and\ \bibinfo {author} {\bibfnamefont
  {D.}~\bibnamefont {Faccio}},\ }\bibfield  {title} {\bibinfo {title} {Imaging
  and certifying high-dimensional entanglement with a single-photon avalanche
  diode camera},\ }\href@noop {} {\bibfield  {journal} {\bibinfo  {journal}
  {npj Quantum Information}\ }\textbf {\bibinfo {volume} {6}},\ \bibinfo
  {pages} {94} (\bibinfo {year} {2020})}\BibitemShut {NoStop}%
\bibitem [{\citenamefont {Jha}\ \emph {et~al.}(2008)\citenamefont {Jha},
  \citenamefont {O'Sullivan}, \citenamefont {Chan},\ and\ \citenamefont
  {Boyd}}]{jha2008pra}%
  \BibitemOpen
  \bibfield  {author} {\bibinfo {author} {\bibfnamefont {A.~K.}\ \bibnamefont
  {Jha}}, \bibinfo {author} {\bibfnamefont {M.~N.}\ \bibnamefont {O'Sullivan}},
  \bibinfo {author} {\bibfnamefont {K.~W.~C.}\ \bibnamefont {Chan}},\ and\
  \bibinfo {author} {\bibfnamefont {R.~W.}\ \bibnamefont {Boyd}},\ }\bibfield
  {title} {\bibinfo {title} {Temporal coherence and indistinguishability in
  two-photon interference effects},\ }\href
  {https://doi.org/10.1103/PhysRevA.77.021801} {\bibfield  {journal} {\bibinfo
  {journal} {Phys. Rev. A}\ }\textbf {\bibinfo {volume} {77}},\ \bibinfo
  {pages} {021801} (\bibinfo {year} {2008})}\BibitemShut {NoStop}%
\bibitem [{\citenamefont {Kulkarni}\ \emph {et~al.}(2017)\citenamefont
  {Kulkarni}, \citenamefont {Kumar},\ and\ \citenamefont
  {Jha}}]{kulkarni2017josab}%
  \BibitemOpen
  \bibfield  {author} {\bibinfo {author} {\bibfnamefont {G.}~\bibnamefont
  {Kulkarni}}, \bibinfo {author} {\bibfnamefont {P.}~\bibnamefont {Kumar}},\
  and\ \bibinfo {author} {\bibfnamefont {A.~K.}\ \bibnamefont {Jha}},\
  }\bibfield  {title} {\bibinfo {title} {Transfer of temporal coherence in
  parametric down-conversion},\ }\href@noop {} {\bibfield  {journal} {\bibinfo
  {journal} {J. Opt. Soc. Am. B}\ }\textbf {\bibinfo {volume} {34}},\ \bibinfo
  {pages} {1637} (\bibinfo {year} {2017})}\BibitemShut {NoStop}%
\bibitem [{\citenamefont {Schneeloch}\ \emph
  {et~al.}(2019{\natexlab{b}})\citenamefont {Schneeloch}, \citenamefont
  {Knarr}, \citenamefont {Bogorin}, \citenamefont {Levangie}, \citenamefont
  {Tison}, \citenamefont {Frank}, \citenamefont {Howland}, \citenamefont
  {Fanto},\ and\ \citenamefont {Alsing}}]{schneeloch2019jopt}%
  \BibitemOpen
  \bibfield  {author} {\bibinfo {author} {\bibfnamefont {J.}~\bibnamefont
  {Schneeloch}}, \bibinfo {author} {\bibfnamefont {S.~H.}\ \bibnamefont
  {Knarr}}, \bibinfo {author} {\bibfnamefont {D.~F.}\ \bibnamefont {Bogorin}},
  \bibinfo {author} {\bibfnamefont {M.~L.}\ \bibnamefont {Levangie}}, \bibinfo
  {author} {\bibfnamefont {C.~C.}\ \bibnamefont {Tison}}, \bibinfo {author}
  {\bibfnamefont {R.}~\bibnamefont {Frank}}, \bibinfo {author} {\bibfnamefont
  {G.~A.}\ \bibnamefont {Howland}}, \bibinfo {author} {\bibfnamefont {M.~L.}\
  \bibnamefont {Fanto}},\ and\ \bibinfo {author} {\bibfnamefont {P.~M.}\
  \bibnamefont {Alsing}},\ }\bibfield  {title} {\bibinfo {title} {Introduction
  to the absolute brightness and number statistics in spontaneous parametric
  down-conversion},\ }\href@noop {} {\bibfield  {journal} {\bibinfo  {journal}
  {Journal of Optics}\ }\textbf {\bibinfo {volume} {21}},\ \bibinfo {pages}
  {043501} (\bibinfo {year} {2019}{\natexlab{b}})}\BibitemShut {NoStop}%
\end{thebibliography}%
	
\end{document}